\documentclass[12pt]{article}
\usepackage{graphicx}
\usepackage{amsmath,amssymb}
\newcommand{\dbox}{\,\raise2pt\hbox{\fbox{\rule{2.5pt}{0pt}\rule{0pt}{2.5pt}}}\,}
\newcommand{\qed}{\,\raise0pt\hbox{\mbox{\rule{6.5pt}{6.5pt}}}}
\newcommand{\bra}[1]{\mbox{$\langle #1 |$}}
\newcommand{\ket}[1]{\mbox{$| #1 \rangle$}}

\begin{document}
\setlength{\baselineskip}{7mm}

\begin{titlepage}
 \begin{normalsize}
  \begin{flushright}
        UT-Komaba/13-10\\
        September 2013
 \end{flushright}
 \end{normalsize}
 \begin{LARGE}
   \vspace{1cm}
   \begin{center}
   Extended string field theory 
   for\\ massless higher-spin fields \\
   \end{center}
 \end{LARGE}
  \vspace{5mm}
 \begin{center}
    Masako {\sc Asano} 
            \hspace{3mm}and\hspace{3mm}
    Mitsuhiro {\sc Kato}$^{\dagger}$ 
\\
      \vspace{4mm}
        {\sl Faculty of Science and Technology}\\
        {\sl Seikei University}\\
        {\sl Musashino-shi, Tokyo 180-8633, Japan}\\
      \vspace{4mm}
        ${}^{\dagger}${\sl Institute of Physics} \\
        {\sl University of Tokyo, Komaba}\\
        {\sl Meguro-ku, Tokyo 153-8902, Japan}\\
      \vspace{1cm}

  ABSTRACT\par
 \end{center}
 \begin{quote}
  \begin{normalsize}
We propose a new gauge field theory which is an extension of ordinary string field theory by assembling multiple state spaces of the bosonic string.
The theory includes higher spin fields in its massless spectrum together with the infinite tower of massive fields.
From the theory, we can easily extract the minimal gauge invariant quadratic action for
tensor fields with any symmetry.
As examples, we explicitly derive the gauge invariant actions for some simple mixed symmetric tensor fields. 
We also construct covariantly gauge-fixed action by extending the method developed for string field theory.

\end{normalsize}
 \end{quote}

\end{titlepage}
\vfil\eject

\section{Introduction}

String theory can be considered as an ultraviolet completion of spin one gauge theory (open string) or spin two gravity (closed string). Then a natural question is what is a UV completion of higher spin gauge theory.  Quantum consistency of the string theory requires the highest spin of the massless mode to be one for open string and two for closed string. Therefore we have to somehow extend the theory in order to adapt it to massless higher spin fields with keeping the consistency of the theory. Constructing such a theory with interaction is not an easy task to complete in every detail. For a free theory level, however, we can systematically construct a class of theories which contain massless higher spin fields as well as massive tower in their spectrum and are invariant under higher spin gauge symmetry.

The purpose of the present paper is to give an extended string-like field theory whose massless mode can have spin higher than two. Our main idea is to consider a multiple tensor product of open string Hilbert spaces as a basis of field space, just like the closed string field basis is a double. For example if we start from $n$ copies of open bosonic string Hilbert space with an appropriate set of conditions, then we obtain a bosonic extended string field theory with massless spin $n$ fields and many other massive fields whose spin $J$ and mass $M$ satisfy the relation \`a la Regge behavior
\begin{equation}J\le \frac{n \ell^2}{2}M^2+n,\end{equation} 
where $\ell$ is a length parameter and is related to the open string slope parameter as $\alpha'=\ell^2/2$. The state density for each mass level is $\rho(M)\sim\exp\big(2\pi n \ell M\sqrt{\frac{D-2}{12}}\,\big)$ in $D$ dimensions, which leads to Hagedorn temperature $T_H\sim\frac{1}{2\pi n \ell}\sqrt{\frac{12}{D-2}}$ in this model with $D=26$. This suggests that the model may have a natural UV cutoff of order $1/\ell$ in the same way as string theory. To clarify this point more precisely, however, we have to incorporate with the interaction, which will be our next task. In the present paper, we only focus on the free theory as a first step.

The model is also useful for extracting the actions for higher spin gauge fields.
We can construct not only gauge invariant free action but also gauge fixed free action systematically utilizing the method developed previously for the open and closed string field theories. The minimal set of fields which describe a given representation of higher rank tensor is systematically identified in the extended string field level.

This paper is organized as follows. In the next section we construct an extended string field theory upon the state space of $n$-tensored version of those for open bosonic string field theory. Both gauge invariant and covariantly gauge-fixed quadratic actions are given there in an analogous way to the string field theory~\cite{Asano:2006hk,Asano:2008iu}. Also minimal gauge invariant set of fields~\cite{Asano:2012qn} are given systematically in the extended string field level. In section~3, we will focus on the massless sector of the model in detail. And some example higher spin gauge field actions extracted from the general action are shown in section~4. The final section will be devoted to the discussions. 
In Appendix~A and B, we collect basic properties of the state space and define  projection operators which are necessary for understanding the gauge fixing and the actions.

\ 

{\it Note added\/}: In the course of writing the manuscript, we became aware of a pioneering paper~\cite{Labastida:1987du} in which the ideas overlapping with us were used. They construct the free gauge fixed action in Feynman-Siegel gauge, while we construct the one in more general covariant gauge including Feynman-Siegel gauge as well as Landau gauge. Also we give a systematic construction of the minimal action, which largely simplifies procedures of extracting a necessary set of component fields for a given higher-spin representation.
\section{Extended string field theory}
In this section, we formulate the new extended string field theory
by assembling $n$ copies of state space of bosonic open string and give a class of gauge fixed actions.
We also give the minimal action which is obtained by eliminating the auxiliary fields part from the original action.
\subsection{State space and inner product}
We first provide $n$ copies of open string state space for momentum $p$ 
and take the direct product of them as 
\begin{equation}
{\cal H}_{n}(p) = {\cal H}^{(1)}(p) \otimes \cdots \otimes{\cal H}^{(n)}(p) .
\end{equation}
Here each ${\cal H}^{(i)}(p)$ ($i=1,\cdots, n$) consists of   
$\alpha_{-l}^{\mu\,(i)}$ $(l\ge 1)$, $c_{-m}^{(i)}$ $(m\ge 0)$, and $b_{-n}^{(i)}$ $(n\ge 1)$ operated on the state 
\begin{equation}
\ket{\downarrow^{i}, 0, p}= \ket{0,p}^{(i)}_{\rm M} \otimes (c_1^{(i)}\ket{0}^{(i)}_{\rm g}) 
\;(= c_1^{(i)} \ket{0,p}^{(i)} )
.
\label{eq:Mggroundsti}
\end{equation}
The state $\ket{0,p}_{\rm M}^{(i)}$ denotes the matter ground state of momentum $p$ 
and $\ket{0}_{\rm g}^{(i)}$ the ghost conformal vacuum
 respectively.
Note that the detailed basic properties of open string state space ${\cal H}(p)$
are collected in \ref{app0}.
We then restrict the space ${\cal H}_{n}(p)$ by imposing the conditions
on $\ket{f}_n \in {\cal H}_n $ as
\begin{equation}
(b_0^{(i)} -b_0^{(j)}  ) \ket{f}_n=0, \qquad (L_0^{(i)} -L_0^{(j)}  ) \ket{f}_n=0 
\label{eq:bLcond}
\end{equation}
for any $i,\,j=1,\cdots, n$.
The operator $L_0^{(i)}$ is given by 
\begin{equation}
L_0^{(i)}= \frac{\ell^2}{2} p^2+ N^{(i)} - 1, 
\qquad\quad \left(\alpha_0^\mu=\ell p^\mu \right)
\label{eq:L0}
\end{equation}
where $N^{(i)}$ counts the level of the ${\cal H}^{(i)}$ part of the state $\ket{f}_n$ .
The resulting restricted space, which we describe as ${\cal H}'_{n}$, is the Hilbert space on which we define our extended string field theory.
Note that the space ${\cal H}'_{n}$ for $n=1$ and $n=2$ correspond to the 
Hilbert spaces for open and closed string field theory respectively.
Also in these cases, the scale parameter $\ell$ introduced in eq.(\ref{eq:L0}) is related to the slope factor as
$\ell=\sqrt{2\alpha'}$ for $n=1$ and $\ell=\sqrt{\alpha'/2}$ for $n=2$.
From the conditions (\ref{eq:bLcond}), 
the operation of $b_0^{(i)}$ and $L_0^{(i)}$ (and thus $N^{(i)}$) on a state in ${\cal H}'_{n}$ gives the equal results independently of $i$.
Thus, we will use the abbreviated notations  
\begin{equation}
b_0^{(i)}=b_0,\qquad  L_0^{(i)}=L_0,\qquad N^{(i)}=N
\label{eq:bLN0}
\end{equation}
on the space ${\cal H}'_{n}$.
Note that each ${\cal H}^{(i)}$ part of any $\ket{f}_n\in {\cal H}_n$ (or ${\cal H}'_{n}$) has the same level $N$ and we call this $N$ as the level of the state $\ket{f}_n$.

As for the usual open or closed string Hilbert space, we can define the  
non-degenerate inner product for states $\ket{f}_n  , \ket{g}_n  \in  {\cal H}'_{n}$
by 
\begin{equation}
\left\langle \ket{f}_n  , \ket{g}_n \right\rangle 
\equiv i^{-n(n-1)/2} \langle {\rm bpz}(f_n) | c_0^{(1)}c_0^{(2)}\cdots c_0^{(n-1)} 
\ket{g}_n 
\label{eq:clinnpro} 
\end{equation}
with the relation 
\begin{equation}
 \langle 0,p| c_{-1}^{(1)}\cdots c_{-1}^{(n)} c_0^{(1)}\cdots c_0^{(n)} c_1^{(1)} \cdots {c}_1^{(n)} \ket{0,p'} =
i^{n(n-1)/2} (2\pi)^D\delta^{(D)}(p-p').
\label{eq:innerpro}
\end{equation}
Here, 
$\langle\mbox{bpz}(f_n)|$ $\equiv$ bpz($\ket{f}_n$) denotes the BPZ conjugate of $\ket{f}_n$. It is related to the Hermitian conjugate $\bra{f} $ of $\ket{f}$ by bpz($\ket{f(p)}$)$=\epsilon_f\bra{f(-p)}$ with $\epsilon_f=\pm 1$. 
The sign of $\epsilon_f$ is explicitly determined by the rules of the BPZ conjugation:  
${\rm bpz}(\ket{0}) = \bra{0}$,
${\rm bpz}(b_{-n}) = (-1)^n b_{n}$,
${\rm bpz}(c_{-n}) = (-1)^{n-1} c_{n}$,
${\rm bpz}(\alpha^{\mu}_{-n}) = (-1)^{n-1} \alpha^{\mu}_{n}$,
and 
${\rm bpz}\,(\alpha \beta) = (-1)^{|\alpha| |\beta|}  {\rm bpz}(\beta) \, {\rm bpz}(\alpha) $
where $|\alpha|$ and $|\beta|$ are the Grassmann parity of $\alpha$ and $\beta$. 
Also, 
\begin{equation}
\ket{0,p}= \ket{0,p}^{(1)} \otimes \cdots \otimes\ket{0,p}^{(n)} 
\end{equation}
where each $\ket{0,p}^{(i)}$ is given by eq.(\ref{eq:Mggroundsti}).
The coefficient $i^{n(n-1)/2}$ in front of the right-hand side of eq.(\ref{eq:innerpro}) is needed for consistency with the hermitian property of the left-hand side since 
\begin{eqnarray}
&&\hspace{-10mm}\left(\langle 0,p| c_{-1}^{(1)}\cdots c_{-1}^{(n)} c_0^{(1)}\cdots c_0^{(n)} c_1^{(1)} \cdots {c}_1^{(n)} \ket{0,p'}\right)^{\mbox{$\ast$}} 
\nonumber\\
&&=
 (-1)^{3n(n-1)/2} \langle 0,p'| c_{-1}^{(1)}\cdots c_{-1}^{(n)} c_0^{(1)}\cdots c_0^{(n)} c_1^{(1)} \cdots {c}_1^{(n)} \ket{0,p} .
\end{eqnarray}

Note that the space ${\cal H}'_{n}$ is divided into two spaces $\tilde{\cal F}_{n}$ and $\tilde{c}_0 \tilde{\cal F}_{n}$
as 
\begin{equation}
{\cal H}'_{n} = \tilde{\cal F}_{n} + \tilde{c}_0 \tilde{\cal F}_{n}
\label{eq:HtildeF}
\end{equation}
where 
\begin{equation}
\tilde{c}_0 = \sum_{i=1}^n c_0^{(i)}
\end{equation}
and 
$\tilde{\cal F}_{n}$ is the set of states of the form 
\begin{equation}
\ket{\tilde{f}}_n = \ket{f^{(1)}}  \otimes \cdots \otimes \ket{f^{(n)} }
\;\in\; b_0^{(1)}c_0^{(1)}{\cal H}^{(1)} \otimes 
\cdots \otimes b_0^{(n)}c_0^{(n)}{\cal H}^{(n)}  .
\label{eq:restf}
\end{equation}
For two states in $\tilde{\cal F}_{n}$ written as
\begin{equation}
\ket{\tilde{f}}_n = {\cal O}_f\ket{\downarrow^1\downarrow^2 \cdots \downarrow^n ;0,p} , 
\qquad
\ket{\tilde{g}}_n= {\cal O}_g\ket{\downarrow^1\downarrow^2 \cdots \downarrow^n ;0,p'}
,
\end{equation} 
the following relation 
\begin{eqnarray}
\left\langle \ket{\tilde{f}}_n  , \tilde{c}_0\ket{\tilde{g}}_n \right\rangle 
&=& i^{-n(n-1)/2} \langle 0,-p| {\rm bpz}({\cal O}_f){\cal O}_g
c_{-1}^{(1)}\cdots c_{-1}^{(n)} c_0^{(1)}\cdots c_0^{(n)} c_1^{(1)} \cdots {c}_1^{(n)} \ket{0,p'}
\nonumber\\
& \equiv &
\bra{0,-p}|  {\rm bpz}({\cal O}_f){\cal O}_g |\ket{0,p'}
  \label{eq:innproF}
\end{eqnarray}
is satisfied.
Here 
\begin{equation}
 \ket{\downarrow^1\downarrow^2 \cdots \downarrow^n ;0,p} 
\equiv \ket{\downarrow^1 ;0,p} \otimes \cdots \otimes 
\ket{\downarrow^n ;0,p}.
\end{equation}
The relation eq.(\ref{eq:innproF}) is useful for dealing with the action given in the following sections since the right-hand side of this relation can be calculated straightforwardly 
by the relation 
\begin{equation}
\bra{0,-p}|1|\ket{0,p'} = (2\pi)^D \delta^{(D)}(p+p')
\end{equation}
derived from eq.(\ref{eq:innerpro}) with the commutation relation of the oscillators.


\bigskip

On the space ${\cal H}'_{n}$, we also define the extended BRST operator $Q_n$ as the sum
of $n$ corresponding open string operators $Q^{(i)}$ $(i=1,\cdots,n)$: $Q_n=\sum_{i=1}^nQ^{(i)}$.
Each $Q^{(i)}$ is divided by ghost zero modes as $Q^{(i)} = \tilde{Q}^{(i)} + c_0^{(i)} L_{0}^{(i)} + b_0^{(i)} M^{(i)} $.
Thus, $Q_n$ can be written on ${\cal H}'_{n}$ as
\begin{equation}
Q_n = \tilde{\cal Q} + \tilde{c}_0 L_0 +b_0 {\cal M}
\end{equation}
with
\begin{equation}
\tilde{\cal Q} = \sum_{i=1}^n \tilde{Q}^{(i)}, \quad
{\cal M} = \sum_{i=1}^n M^{(i)}
=\sum_{i=1}^n \left(-\sum_{k=1}^\infty 2k c_{-k}^{(i)} c_k^{(i)}\right)
\end{equation}
where
 $L_0$ and $b_0$ on ${\cal H}'_{n}$ are respectively given by $L_0^{(i)}$ and $b_0^{(i)}$ for any $i$ as we have mentioned in eq.(\ref{eq:bLN0}).
We also define the following operators 
\begin{equation}
{\cal M}^-= \sum_{i=1}^n M^{-(i)}
=-\sum_{i=1}^n \sum_{k=1}^\infty \frac{1}{2k} b_{-k}^{(i)} b_k^{(i)}
,\qquad
{\cal M}_z = \sum_{i=1}^n \sum_{k=1}^\infty \frac{1}{2}
\left(c_{-k}^{(i)} b_k^{(i)}- b_{-k}^{(i)} c_k^{(i)}\right)
\end{equation}
on ${\cal H}_n$.
Here, ${\cal M}_z$ is the operator which counts the half of the ghost number of non-zero mode part: ${\cal M}_z= \frac{1}{2}\hat{G}$, and the relation to the ghost number operator is given by $G=\hat{G}+ \tilde{c}_{0} b_0+n$ on ${\cal H}'_{n}$.
Note that the structure of the space ${\cal H}'_{n}$ (or $\tilde{\cal F}_{n}$ and  $\tilde{c}_0\tilde{\cal F}_{n}$) under the operators $\tilde{\cal Q}$, $L_0$ ${\cal M}$,
${\cal M}^-$ and ${\cal M}_z$ is determined independently of $n$ since the commutation relations among 
the operators are the same as for $n=1$ or $n=2$ cases explained in refs.\cite{Asano:2006hk,Asano:2008iu,Asano:2012sk}.
For example, ${\cal M}$, ${\cal M}^-$ and ${\cal M}_z$ constitute the SU(1,1) algebra 
with the relation 
\begin{equation}
 [{\cal M},{\cal M}^-] = 2 {\cal M}_z ,
\quad [{\cal M}_z,{\cal M}] = {\cal M}, 
\quad [{\cal M}_z,{\cal M}^-] = -{\cal M}^-.  
\end{equation}
Furthermore, if we divide the space $\tilde{\cal F}_{n}$ by the non-zero mode part of the ghost number $\hat{G}$ as 
\begin{equation}
\tilde{\cal F}_{n} =\bigoplus_{ \hat{g} =-\infty}^\infty {\cal F}_n^{\hat{G}= \hat{g} },
\end{equation}
${\cal F}_n^{\hat{g}}$ and  ${\cal F}_n^{-\hat{g}}$ are isomorphic to each other and 
we can define the operator ${\cal W}_{\hat{g}} $ which is considered as the inverse of 
${\cal M}^{\hat{g}}$ on the space ${\cal F}_n^{\pm \hat{g}}$:
For any $\ket{\tilde{f}^{\hat{g}} } \in \tilde{\cal F}_n^{\hat{g}}$
or $\ket{\tilde{f}^{-\hat{g}} } \in \tilde{\cal F}_n^{-\hat{g}}$ with $\hat{g}>0$,
\begin{equation}
 {\cal M}^{\hat{g}} {\cal W}_{\hat{g}} \ket{\tilde{f}^{\hat{g}} } =\ket{\tilde{f}^{\hat{g}} },
\qquad 
{\cal W}_{\hat{g}} {\cal M}^{\hat{g}} \ket{\tilde{f}^{-\hat{g}} } =\ket{\tilde{f}^{-\hat{g}} }.
\end{equation}
Explicitly, ${\cal W}_{\hat{g}}$ is represented by using ${\cal M}$ and ${\cal M}^-$ as
\begin{equation}
 {\cal W}_{\hat{g}} = 
\sum_{i=0}^{\infty}\, (-1)^i \frac{(\hat{g}+i-1)!}{[(\hat{g}+i)!]^2\, i!\,(\hat{g}-1)!}\,
({\cal M}^-)^{\hat{g}+i} {\cal M}^i .
\label{eq:defWg}
\end{equation}

\subsection{Gauge invariant and gauge fixed action}
We have prepared the state space for our extended string field theory.
Now we define the free action for any $n$ 
by extending the string field theory action as
\begin{equation}
S_n=- \frac{1}{2}\langle \Phi_n , Q_n \Phi_n \rangle .
\label{eq:action2}
\end{equation}
Here, $\Phi_n$ is the extended `string field' 
which is expanded by states $\ket{\tilde{f}^k(p)} \in {\cal H}'_n (p)$ 
with the corresponding coefficient fields $\psi_{\tilde{f}^k}(p)$ as
\begin{equation}
\Phi_n = \int \frac{d^{D}p }{(2\pi)^D}\sum_k \ket{\tilde{f}^k(p)} \,\psi_{\tilde{f}^k}(p),
\qquad  \ket{\tilde{f}^k(p)} \in {\cal H}'_n (p) .
\label{eq:expandstatefield2}
\end{equation}
Note that $\Phi_n$ has ghost number $G=n$ and the Grassmann parity $(-)^{n}$.
Also, according to the division of the state space eq.(\ref{eq:HtildeF}), 
$\Phi_n$ is divided into two parts $\Phi_n= \phi_{n} +\tilde{c}_0 \omega_{n}$
where $\phi_n=b_0\tilde{c}_0 \Phi_n$ and $\omega_{n}= b_0 \Phi_n$.
The non-zero mode part ghost number $\hat{G}$ 
of $\phi_{n}$ and $\omega_{n}$ are respectively $0$ and $-1$.

In order that the action $S_n$ is hermitian and has the correct sign factor, we should assign  the appropriate hermiticity for each field $ \psi_{\tilde{f}^k} (p)$:
For a state $\ket{\tilde{f}^k(p)}$ which satisfy 
\begin{equation}
\mbox{bpz}(\ket{\tilde{f}^k(p)})=(-)^{n(n-1)/2}\tilde{\epsilon}_{\tilde{f}^k} \bra{\tilde{f}^k(-p)} ,
\end{equation}
the corresponding field $\psi_{\tilde{f}^k} (p)$ should have the property 
\begin{equation}
(\psi_{\tilde{f}^k} (-p))^* = \tilde{\epsilon}_{\tilde{f}^k} \psi_{\tilde{f}^k} (p).
\end{equation}
Thus, if we fix the field $\psi_{\tilde{f}^k} (p)$ to be real, 
we should associate $i$ in front of the field when $\tilde{\epsilon}_{\tilde{f}^k}=-1 $.

If we take $D=26$, the action eq.(\ref{eq:action2}) is invariant under the gauge transformation
\begin{equation}
\delta \Phi_n = Q_n \Lambda_n
\label{eq:ginv}
\end{equation}
for any $G=n-1$ extended string field $\Lambda_n $ with Grassmann parity $(-)^{n-1}$.

We will see the spectrum of $\Phi_n$ for general $n$ for $D=26$. 
The lowest mass level ($N=0$, $M^2=- 2/\ell^2 $) gives the tachyon field $\phi$
as the coefficient of the state 
\begin{equation}
 \ket{\downarrow^1\downarrow^2 \cdots \downarrow^n ;0,p}.
\end{equation}
The next massless level ($N=1$, $M^2=0$) gives the $n$-th rank tensor field $A_{\mu_1\cdots\mu_n}$ as the coefficient of the state 
\begin{equation}
\alpha_{-1}^{\mu_1(1)} \alpha_{-1}^{\mu_2 (2)}\cdots\alpha_{-1}^{\mu_n (n)}
 \ket{\downarrow^1 \cdots \downarrow^n ;0,p} 
\label{eq:N1stategenA}
 \end{equation}
with the lower rank $n-2k$ ($k=1,2,\cdots$) tensor fields given by the coefficients of the states including $b^{(i)}_{-1}$ and $c^{(i)}_{-1}$ ghost fields. 
Thus we see that the massless spectrum for $n$ includes fields with 
spin up to $n$.
As we will see below, the $N=1$ (and also $N=0$) part of the quadratic action is consistent for any spacetime dimension $D$. 
The number of massless physical degrees of freedom in $D$-dimension is $(D-2)^n$
which can be counted by the little group SO($D-2$). For example, totally symmetric tensor part has $\Big(\!{\scriptsize\begin{array}{c}D-3+n\\n\end{array}}\!\Big)$ degrees of freedom while totally anti-symmetric tensor part does $\Big(\!{\scriptsize \begin{array}{c}D-2\\n\end{array}}\!\Big)$.
We will investigate the structure of the massless part of the action in detail
in the next section.

There are also infinite tower of massive states with mass $M^2 = 2(N-1)/\ell^2$ 
corresponding to the level $N$.
In general, the number of physical degrees of freedom can be counted by the transverse 
oscillators thanks to the equivalence to the light-cone gauge, as is usual in string theory.  
Indeed we can construct the analogue of transverse DDF operators~\cite{Del Giudice:1971fp} which form the spectrum generating algebra.
Also we can extend usual tree-level no-ghost theorem for $n=1,2$ to $n>2$ cases, 
thereby this free theory is consistent quantum mechanically.  

As is the case for open or closed string field theory action, 
the above gauge invariance given in eq.(\ref{eq:ginv}) is precisely fixed by the extension of Feynman-Siegel gauge~\cite{Labastida:1987du,Siegel:1984wx} or $a$-gauges~\cite{Asano:2006hk,Asano:2008iu,Asano:2012sk}, or other appropriate gauges.
General form of the gauge fixed action is given by 
\begin{equation}
S_{n,{\rm GF}}  = 
-\frac{1}{2}  \sum_{m=-\infty}^{\infty} \left\langle \Phi_{n}^{m}, Q_n  \Phi_{n}^{-m+2} \right\rangle
+  \sum_{m=-\infty}^{\infty} 
\left\langle {\cal O}_n^{\langle -m+4 \rangle}  {\cal B}_{n}^{-m+4} , \Phi_{n}^{m}  \right\rangle 
\end{equation}
where $\Phi_n^m$ and ${\cal B}_{n}^{m} $ are extended string fields 
with ghost number $m+n-1$ and the Grassmann parity $(-)^{n}$.
Note that $\Phi_n^1 = \Phi_n$.
For each of the gauge choices, the operators ${\cal O}_n^{\langle m \rangle}$ are given in order that the action $S_{n,{\rm GF}}$ remains no gauge invariance.
In particular, as for the $a$-gauges, since the structure of the BRST operator $Q_n$ with respect to the ghost zero modes is equivalent for every $n$, 
general procedure for constructing the gauge fixed action is given parallel to the case of open or closed string field theory~\cite{Asano:2006hk,Asano:2008iu,Asano:2012sk}
and the operators ${\cal O}_{n,a}^{\langle m \rangle}$ for $a$-gauges are explicitly given by
\begin{eqnarray}
{\cal O}_a^{\langle m+1 \rangle}  &=& \frac{1}{1-a}
(b_0 + a \tilde{c}_0 b_0  \tilde{\cal Q} {\cal M}^{m-2 }  {\cal W}_{m-1} ) \hspace*{3cm} (m\ge 2),
\\
{\cal O}_a^{\langle -m+4 \rangle} &=& b_0 (1-{\cal P}_{-m+2})+
\frac{1}{1-a}(b_0 {\cal P}_{-m+2} +
a \tilde{c}_0 b_0  {\cal W}_{m-1} {\cal M}^{m-2} \tilde{\cal Q} )\quad (m\ge 2)
\end{eqnarray}
where the definition of ${\cal P}_{m} $ is given in \ref{app1}.
Note that if we take $a=0$, ${\cal O}_{a=0}^{\langle m \rangle} =b_0$ which gives the Feynman-Siegel gauge.
Note also that we can extend the $a$-gauges to admit multiple number of gauge fixing parameters as in ref.\cite{Asano:2012qn}.


\subsection{Minimal gauge invariant and gauge fixed action}
As stated in the previous subsection, $\Phi_n$ can be decomposed into two parts $\Phi_n=\phi_n+\tilde c_0\omega_n$. $\phi_n$ has a nontrivial kinetic term while $\omega_n$ is an auxiliary field. In fact, gauge invariant action is also divided into two parts each of which is gauge invariant independently. Thus we can use a minimal set of fields collected into the extended string field $\phi_n$ in order to obtain minimal gauge invariant action as discussed in ref.\cite{Asano:2006hk, Asano:2012qn} for the case of open string field theory.
In concrete, the gauge invariant action $S_n$ is divided into the minimal action part $S_n^{\rm min}$ and the auxiliary field part $S_n^{\rm auxiliary}$.
The former minimal action part is given by
\begin{equation}
S_n^{\rm min} = 
-\frac{1}{2} \left\langle  \phi_n, \tilde{c}_0 (L_0+\tilde{\cal Q}{\cal W}_1  \tilde{\cal Q})\phi_n\right\rangle.
\label{eq:Snmin}
\end{equation}
We may restrict $\phi_n$ to satisfy $M\phi_n=0$ in this action 
since otherwise $(L_0+\tilde{\cal Q}{\cal W}_1  \tilde{\cal Q})\phi_n=0$ in any case.
Note that if we use the projection operator $1-{\cal P}_0$ which extracts the states satisfying $\tilde{\cal Q} \phi_n = 0 $, 
the above action can be represented as
\begin{equation}
S_n^{\rm min} = 
-\frac{1}{2} \left\langle  \phi_n, \tilde{c}_0 L_0(1-{\cal P}_0)\phi_n\right\rangle.
\label{eq:Snmin2}
\end{equation}
The definition of ${\cal P}_0 $ is given in \ref{app1}.

This minimal action $S_n^{\rm min}$ is invariant under the gauge transformation
\begin{equation}
\delta\phi_n = \tilde{\cal Q} \lambda_n
\end{equation}
with
\begin{equation}
\lambda_n = b_0\tilde{c}_0 \Lambda_n.
\end{equation}
We can fix this gauge invariance by taking generalized $a$-gauges and 
construct the corresponding gauge fixed action $S_{n, \{\alpha \}}$ by the procedure 
parallel to the case of open string field theory as we discussed in ref.\cite{Asano:2012qn}.
Here we only present the resulting gauge fixed action as follows.
It is given by the sum of the gauge invariant action $S_n^{\rm min}$ and the (anti-)ghosts plus gauge fixing terms:
\begin{equation}
S_{n, \{\alpha \}} = S_n^{\rm min} + S_{n;\,{\rm gh+gf},\{\alpha\}} 
\label{eq:Salpha}
\end{equation}
where
\begin{eqnarray}
 S_{n;\,{\rm gh+gf},\{\alpha\}} 
 &=& 
-\sum_{m=1}^\infty \langle\phi_n^{(m)}, \tilde{c}_0 L_0(1-{\cal P}_{-m})\phi_n^{(-m)}\rangle
\nonumber\\
&&
+
\sum_{m=0}^\infty \langle \tilde{c}_0 \beta_n^{(m+1)}, {\cal W}_{m+1}{\cal M}^m\tilde{\cal Q}\phi_n^{(-m)}\rangle
+\sum_{m=1}^\infty \langle \tilde{c}_0 \beta_n^{(-m+1)}, {\cal M}^m {\cal W}_{m+1}\tilde{\cal Q}\phi_n^{(m)}\rangle
\nonumber\\
&&
+\sum_{m=0}^\infty 
\,\,\sum_{k\in \{ \frac{m+1}{2} + 
{Z}_{\ge 0}
 \}} 
\!\!\alpha^k_{(-m+1,m+1)} 
\langle {\cal S}_k\beta_n^{(-m+1)}, \tilde{c}_0 {\cal M}^m {\cal W}_{m+1} 
{\cal S}_k\beta_n^{(m+1)}\rangle.
\label{eq:SFPgf}
\end{eqnarray}
Here, ${\cal W}_m$ is given by eq.(\ref{eq:defWg}), and  
${\cal P}_m$ and  ${\cal S}_k$ are certain projection operators defined in \ref{app1}. 
Also, $\phi^{(0)}_n=\phi_n$ and $\phi^{(m)}_n$ and $\beta^{(m)}_n$ are $b_0=0$ part of the ghost number $G=n+m$ extended string fields.
Grassmann parity of $\phi^{(m)}_n$ and $\beta^{(m)}_n$ are 
$(-)^n$ and $(-)^{n+1}$ respectively.
Thus, Grassmann parity of the fields associated with each of extended string states in 
$\phi^{(m)}_n$ and $\beta^{(m)}_n$ are $(-)^m$ and $(-)^{m+1}$ respectively.

In the above action, fields associated with SU(1,1)-spin$=k\,(> 0)$~\cite{Asano:2006hk} part of extended string fields $\phi^{(l)}_n$ and $\beta^{(l')}_n$ plays the role of $2k$-th order (anti-)ghost fields and the Lagrange multiplier fields needed for fixing the corresponding gauge invariance.
Here, the possible $l$ and $l'$ for each $k$ are given respectively as $l=-2k+2i$ with $i=0,1,\cdots,2k$ and $l'=-2k+2j$ with $j=1,2, \cdots,2k$.
Note that $\alpha^k_{(-m+1,m+1)} (= \alpha^k_{(m+1,-m+1)})$ are arbitrary parameters distinguishing the gauge fixing conditions. 
For the $2k$-th order (anti-)ghost field terms, we can choose $[k+\frac{1}{2}]$ number of independent real parameters $\alpha^k_{(-m+1,m+1)}$ with $m=2k-1,2k-3,\cdots,1$ (or $m=2k-1,2k-3,\cdots,0$) for an integer $k$ (or a half integer $k$). 
Here, $[x]$ is the floor function which gives the largest integer not greater than $x$.
If we choose all the $\alpha^k_{(-m+1,m+1)}$ to be $1$, the action becomes the one for the Feynman-Siegel gauge.

For each gauge, we can calculate the propagators from the gauge fixed action eq.(\ref{eq:Salpha}) in a similar way for the $n=1$ open string field theory case~\cite{Asano:2012sk}. 
Note also that the action given in eq.(\ref{eq:Salpha}) is invariant under the BRST and the anti-BRST transformations $\delta_{\rm B}$ and $\tilde{\delta}_{\rm B}$ satisfying the condition 
\begin{equation}
\delta_{\rm B}{}^2= \tilde{\delta}_{\rm B}{}^2 = \{\delta_{\rm B},\tilde{\delta}_{\rm B}\}=0. 
\end{equation}
For the detailed properties of the action eq.(\ref{eq:Salpha}), consult ref.\cite{Asano:2012sk}.

\section{Massless part of the minimal action}
In this section, we in particular consider the massless part of the minimal action 
$S_n^{{\rm min}, N=1}$ 
since it is
a useful tool for obtaining simple gauge invariant actions for general higher spin fields expressed by higher rank tensor fields with various types of symmetry in a systematic manner.
Furthremore, if we limit ourselves to the massless level $N=1$ fields, 
the action is consistent not only for the spacetime dimensions $D=26$, but for any $D$.

The operators $\tilde{\cal Q}$, ${\cal M}$, ${\cal M}^-$ 
and ${\cal M}_z$ for $N=1$ are reduced to 
\begin{equation}
\tilde{\cal Q}_{N\!=\!1} = \sum_{i=1}^n \ell p_\mu 
\left( \alpha_{-1}^{\mu (i)}c_1^{(i)}  +   c_{-1}^{(i)} \alpha_{1}^{\mu (i)}\right),
\end{equation}
\begin{equation}
{\cal M}_{N\!=\!1} = -2 \sum_{i=1}^n  c_{-1}^{(i)} c_{1}^{(i)} ,
\qquad 
{\cal M}_{N\!=\!1}^- = -\frac{1}{2} \sum_{i=1}^n  b_{-1}^{(i)} b_{1}^{(i)},
\end{equation}
and
\begin{equation}
 {\cal M}_{z,N\!=\!1} \left(=\frac{1}{2} \hat{G}_{N\!=\!1}\right)
 = \frac{1}{2} \sum_{i=1}^n 
 \left( c_{-1}^{(i)} b_{1}^{(i)} - b_{-1}^{(i)} c_{1}^{(i)}  \right).
\end{equation}
Extended string field ${\phi}_n^{N=1}$ for $N=1$ consists only of 
$\alpha^{(i)}_{-1}$, $b^{(i)}_{-1}$ and $c^{(i)}_{-1}$. 
Thus, the general form of ${\phi}_n^{N=1}$ is given by 
\begin{equation}
{\phi}_n^{N\!=\!1} =\int \frac{dp^D}{(2\pi)^D} \tilde{\phi}_n^{N\!=\!1}(p)
\end{equation}
with 
\begin{eqnarray}
\tilde{\phi}_n^{N=1}(p) &=& 
\alpha_{-1}^{\mu_1(1)} \alpha_{-1}^{\mu_2 (2)}\cdots\alpha_{-1}^{\mu_n (n)}
 \ket{\downarrow^1 \cdots \downarrow^n ;0,p} 
A_{\mu_1\mu_2\cdots\mu_n} (p)
\nonumber\\
&+&
\sum_{i_1,{j}_1}
\alpha_{-1}^{\mu_1 (k_1)} \cdots\alpha_{-1}^{\mu_{n-2} (k_{n-2})}
c_{-1}^{(i_1)}b_{-1}^{(j_1)}
 \ket{\downarrow^1 \cdots \downarrow^n ;0,p} 
A_{\mu_1\cdots\mu_{n-2}}^{i_1;{j}_1} (p)
\quad+\cdots 
\nonumber\\
&+&\!\!\!\!\!\!\!
\sum_{i_1< i_2<\cdots < i_q ;j_1< j_2<\cdots < j_q}\!\!\!\!\!\!\!
\alpha_{-1}^{\mu_1 (k_1)} \cdots\alpha_{-1}^{\mu_{n-2q} (k_{n-2q})}
c_{-1}^{(i_1)}\cdots c_{-1}^{(i_q)} b_{-1}^{(j_1)}\cdots b_{-1}^{(j_q)} 
 \ket{\downarrow^1 \cdots \downarrow^n ;0,p} 
 \nonumber\\[-3pt]
&& \hspace*{10cm}
\times A_{\mu_1\cdots\mu_{n-2q}}^{i_1,\cdots,i_q ;j_1,\cdots,j_q} (p)
\nonumber\\[3pt]
&+& \cdots 
\label{eq:N1stategen}
\end{eqnarray}
where the summation of indices $i_m$ and $j_m$ ($1\le m \le q$ for each $q$) are taken from $1$ to $n$ under the condition that all those indices are different values.
For each set of $i_m$ and $j_m$, the indices $k_l$ ($1\le l\le n-2q$)
are uniquely determined by the conditions $k_l\ne i_m$, $k_l\ne j_m$ 
and $1\le k_1<k_2 \cdots < k_{n-2q}\le n$. 
Thus, for any choice of $i_m$ and $j_m$ for each $q$, the set of $k_l$, $i_m$ and $j_m$ 
coincides with the set of $n$ integers $1,\cdots, n$, {\it i.e.},
\begin{equation}
\{k_1, \cdots, k_{n-2q}, i_1,\cdots,i_q, j_1,\cdots,j_q \}=\{1,2,\cdots,n\}.
\end{equation}
We see that there appear tensors of rank $r=n$, $n-2$, $\cdots$, $n-2q$, $\cdots$
in $\phi_n^{N=1}$, and the lowest rank tensors are scalars ($r=0$) for even $n$ or 
vectors ($r=1$) for odd $n$.
As we have already mentioned in the previous section,
among the above general form of $\phi_n^{N=1}$,
only the terms that satisfy the condition $M \phi_n^{N=1} = 0$
appear in the action.
For example, the $M \phi_n^{N=1} = 0$ part of the terms of rank $r=n-2$  tensors in eq.(\ref{eq:N1stategen}) is
given by
\begin{equation}
\sum_{i_1<{j}_1}
\alpha_{-1}^{\mu_1 (k_1)} \cdots\alpha_{-1}^{\mu_{n-2} (k_{n-2})}
(c_{-1}^{(i_1)}b_{-1}^{(j_1)}- b_{-1}^{(i_1)}c_{-1}^{(j_1)})
 \ket{\downarrow^1 \cdots \downarrow^n ;0,p} 
A_{\mu_1\cdots\mu_{n-2}}^{i_1;{j}_1} (p).
\end{equation}

The gauge invariant action for $N=1$ is given by 
\begin{eqnarray}
S^{{\rm min},N=1}_n &=&
-\frac{1}{2} \left\langle  \phi_n^{N=1}, \tilde{c}_0 
\left(
\frac{\ell^2 p^2}{2} + \tilde{\cal Q}_{N=1} {\cal W}_{1} \tilde{\cal Q}_{N=1}
\right)
\phi_n^{N=1}
\right\rangle
\label{eq:N1action}
\\
 \Big( &=& \int{dx^D {\cal L}_n^{{\rm min},N=1 }  } \Big).
\end{eqnarray}
Unlike the general action, the above massless part of the action 
$S^{{\rm min},N=1}_n$
 is invariant under the gauge transformation 
\begin{equation}
\delta\phi_n^{N=1} = \tilde{\cal Q}_{N=1} \lambda_n^{N=1}
\label{eq:N1gaugetr}
\end{equation}
not only for the spacetime dimension $D=26$, but also for arbitrary $D$
since the relation $\tilde{\cal Q}_{N=1}^2 = - \frac{\ell^2 p^2}{2}{\cal M}_{N=1} $ is satisfied for general $D$.
General form of the gauge parameter field is given by
\begin{equation}
\lambda_n^{N=1} =\int \frac{dp^D}{(2\pi)^D} \tilde{\lambda}_n^{N=1}(p)
\end{equation}
with 
\begin{eqnarray}
\tilde{\lambda}_n^{N=1}(p) &=& 
\sum_{j=1}^n
\alpha_{-1}^{\mu_1 (k_1)} \cdots\alpha_{-1}^{\mu_{n-1} (k_{n-1})}
b_{-1}^{(j)}
 \ket{\downarrow^1 \cdots \downarrow^n ;0,p} 
\lambda_{\mu_1\cdots\mu_{n-1}}^{j} (p)
\quad+\cdots 
\nonumber\\
&+&
\!\!\!\!\!\!\!\!\!\!\!\!\!\!\!
\sum_{i_1< i_2<\cdots < i_q ; j < j_1< j_2<\cdots < j_q}\!\!\!\!\!\!\!\!\!\!\!
\alpha_{-1}^{\mu_1 (k_1)} \cdots\alpha_{-1}^{\mu_{n-2q-1} (k_{n-2q-1})}
c_{-1}^{(i_1)}\cdots c_{-1}^{(i_q)} b_{-1}^{(j)}b_{-1}^{(j_1)}\cdots b_{-1}^{(j_q)} 
 \ket{\downarrow^1 \cdots \downarrow^n ;0,p} 
 \nonumber\\[-3pt]
&& \hspace*{10cm}
\times \lambda_{\mu_1\cdots\mu_{n-2q-1}}^{i_1,\cdots,i_q ;j, j_1,\cdots,j_q} (p)
\nonumber\\[3pt]
&+& \cdots 
\end{eqnarray}
where as in the case of eq.(\ref{eq:N1stategen}), the summation of indices $j$, $i_m$, and $j_m$ are taken from $1$ to $n$ under the condition that all those indices are different values for each $q$.
Then the values $k_l$ ($1\le l \le n-2q-1 $) are determined so as to satisfy
$k_1<k_2 \cdots < k_{n-2q-1}$ and
\begin{equation}
\{k_1, \cdots, k_{n-2q-1}, i_1,\cdots,i_q, j, j_1,\cdots,j_q \}=\{1,2,\cdots,n\}
\end{equation}
for each integer $q$ satisfying $q\ge 0$ and $n-2q-1 \ge 0$.
Note that only the SU(1,1)-spin $=1/2$ part of $\lambda_n^{N=1}$ appears in the gauge transformation eq.(\ref{eq:N1gaugetr}).
The gauge fixed action for $S^{{\rm min},N=1}_n$ is extracted from $N=1$ part 
of $S_{n, \{\alpha \}}$ given in eq.(\ref{eq:Salpha}) with eq.(\ref{eq:SFPgf}).

For each $n$, the extended string field $\phi_n^{N=1}$ and the gauge parameter field $\lambda_n^{N=1}$ are divided into several independent parts by the symmetry of the action (\ref{eq:N1action}) and the gauge transformation (\ref{eq:N1gaugetr}).
For example, for $n=2$, fields in ${\phi}_{n=2}^{N=1}$ 
and correspondingly the action are divided into two parts as follows: 
First, we divide the second rank tensor $A_{\mu\nu}$ part of ${\phi}_{n=2}^{N=1}$ into the symmetric tensor $A_{(\mu\nu)}$ part and the anti-symmetric tensor $A_{[\mu\nu]}$ part.
The basis states for $A_{(\mu\nu)}(p)$ and $A_{[\mu\nu]}(p)$ are respectively given by
\begin{equation}
(\alpha_{-1}^{\mu (1)}\alpha_{-1}^{\nu (2)} \pm
\alpha_{-1}^{\nu (1)}\alpha_{-1}^{\mu (2)} 
)\ket{\downarrow\downarrow;0,p} .
\end{equation}
Accordingly, other fields in ${\phi}_{n=2}^{N=1}$ and ${\lambda}_{n=2}^{N=1}$, and thus the action is divided into the two parts.
Similarly, for general $n$, classification of the fields and the action 
is given by the symmetry of the $n$-th rank tensor $A_{\mu_1\mu_2\cdots\mu_n}$, which is explicitly specified by the Young diagrams of $n$ boxes.
For each Young diagram, we are able to choose an appropriate state as 
the basis of the corresponding tensor field of the symmetry.
Such a basis state in general has the form 
\begin{equation}
\sum a_{i_1,\cdots,i_n} \, \alpha_{-1}^{\mu_1 (i_1)}\cdots \alpha_{-1}^{\mu_n (i_n)}
\ket{\downarrow\cdots \downarrow;0,p}  
\label{eq:basisstn}
\end{equation}
where the coefficient $a_{i_1,\cdots,i_n}$ is determined by the symmetry of the corresponding Young diagram.
Note that there are in general several possible choices of the coefficient $a$ for mixed symmetric cases.
Explicitly, the number of independent bases for a particular diagram is given by
$n!/[\prod_{i=1}^n \mbox{hook}(x_i) ]$ where hook$(x_i)$ denotes the hook length of the box $x_i$.
Once an explicit basis is chosen, we can construct the gauge invariant action for the corresponding $n$-th rank tensor field by extracting the relevant part from the general Lagrangian ${\cal L}_n^{{\rm min},N=1}$.
As a result, all the $\phi_n^{N=1}$, $\lambda_n^{N=1}$ and ${\cal L}_n^{{\rm min},N=1}$ are divided by symmetry as
\begin{equation}
\phi_n^{N=1}=\sum_{s} \phi_n^{N=1,(s)},
\qquad
\lambda_n^{N=1}=\sum_{s} \lambda_n^{N=1,(s)}
\end{equation}
and
\begin{equation}
{\cal L}_n^{{\rm min},N=1}= \sum_{s} {\cal L}_n^{N=1,(s)}(\phi_n^{N=1,(s)} )
\end{equation}
where $s$ denotes the index specifying the independent part of the action 
divided by the symmetry of the $n$-th rank tensor field represented by the Young diagram. 
Each $\phi_n^{N=1,(s)}$ includes one $n$-th rank tensor of a particular symmetry with some associated lower rank tensor fields in general.
In this way, we can explicitly construct the gauge invariant action for $n$-th rank tensor field of arbitrary symmetry.

\section{Examples of massless gauge invariant action}
In this section, we will explicitly give some examples of gauge invariant actions for several simple Young diagrams following the procedures given in the previous section.
In the following, we set the scale factor as $\ell=1$  
and represent the oscillators in the abbreviated forms as 
$\alpha_{-1}^{\mu (i)}\rightarrow \alpha_{(i)}^\mu$,
$b_{-1}^{(i)}\rightarrow b_{(i)}$ and $c_{-1}^{(i)}\rightarrow c_{(i)}$
for simplicity.

\subsection{$n=3$ action}
First, we consider the $n=3$ theory.
In this case, fields are given by eq.(\ref{eq:N1stategen}) for $n=3$ with the condition $M=0$ and thus we have a general third rank tensor field $A_{\mu\nu\rho}$ and three vector fields $D_{\mu}^i$ ($i=1,2,3$) in the extended string field as
\begin{eqnarray}
\tilde{\phi}_{n=3}(p) &=&
\alpha_{(1)}^{\mu} \alpha_{(2)}^{\nu}\alpha_{(3)}^{\rho}
 \ket{\downarrow\downarrow\downarrow ;0,p} 
A_{\mu\nu\rho} (p)
\nonumber\\
&&+
\sum_{i=1}^3
\alpha_{(i)}^{\mu} (c_{(i+1)} b_{(i+2)} -  b_{(i+1)} c_{(i+2)}   )
 \ket{\downarrow\downarrow\downarrow  ;0,p} 
D_{\mu}^{i} (p).
\label{eq:phiADn3}
\end{eqnarray}
Here and in the remainder of this subsection, the indices in parentheses including the character $i$ are to be interpreted as mod 3 numbers.

\begin{figure}[ht]
  \centering
\setlength{\unitlength}{1.2pt}
\begin{picture}(130,60)(0,0)
\put(-20,20){\framebox(30,10){}}
\put(-10,20){\line(0,1){10}}
\put(0,20){\line(0,1){10}}
\put(-30,40){{(i)}}
\put(60,40){{(ii)}}
\put(70,10){\line(1,0){10}}
\put(70,10){\line(0,1){20}}
\put(80,10){\line(0,1){20}}
\put(70,20){\line(1,0){20}}
\put(70,30){\line(1,0){20}}
\put(90,20){\line(0,1){10}}
\put(140,40){{(iii)}}
\put(150,0){\framebox(10,30){}}
\put(150,10){\line(1,0){10}}
\put(150,20){\line(1,0){10}}
\end{picture}
  \caption{Three types of Young diagrams for $n=3$}
  \label{fig:n3Young}
\end{figure}
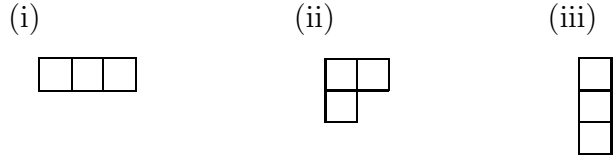
Since there are three kinds of Young diagrams for $n=3$
as depicted in fig.\ref{fig:n3Young}, the general third rank tensor field $A_{\mu\nu\rho}$ is divided into three parts by symmetry.
In fact, the third rank tensor part of $\tilde{\phi}_{n=3}$ is divided into 
four independent parts since there are two independent bases for the diagram (ii) of fig.\ref{fig:n3Young}. 
This is explicitly given by
\begin{eqnarray}
\tilde{\phi}_{n=3}^{N=1}|_{A_{\mu\nu\rho}}
&=&
\alpha_{(1)}^{(\mu} \alpha_{(2)}^{\nu}\alpha_{(3)}^{\rho)}
\ket{\downarrow \downarrow \downarrow ;0,p} 
A^{\mbox{\scriptsize (i)}}_{(\mu\nu\rho)}(p) \qquad \qquad\qquad\qquad\quad\;\,\cdots \mbox{(i)}
\\
& +&
\left(
\alpha_{(1)}^{(\mu} \alpha_{(2)}^{\nu)}\alpha_{(3)}^{\rho}
-\alpha_{(1)}^{\rho } \alpha_{(2)}^{(\mu}\alpha_{(3)}^{\nu)}
\right)
\ket{\downarrow \downarrow \downarrow ;0,p} 
A^{\mbox{\scriptsize (ii-1)}}_{(\mu\nu),\rho}(p) \quad\;\;\,\cdots \mbox{(ii-1)}
\\
& +&
\left(
\alpha_{(1)}^{[\mu} \alpha_{(2)}^{\nu]}\alpha_{(3)}^{\rho}
-\alpha_{(1)}^{\rho } \alpha_{(2)}^{[\mu}\alpha_{(3)}^{\nu]}
\right)
\ket{\downarrow \downarrow \downarrow ;0,p} 
A^{\mbox{\scriptsize (ii-2)}}_{[\mu\nu],\rho}(p) \qquad \cdots \mbox{(ii-2)}
\\
& +&
\alpha_{(1)}^{[\mu} \alpha_{(2)}^{\nu}\alpha_{(3)}^{\rho]}
\ket{\downarrow \downarrow \downarrow ;0,p} 
A^{\mbox{\scriptsize (iii)}}_{[\mu\nu\rho]}(p) .\qquad\qquad\qquad\qquad\quad\; \,\cdots \mbox{(iii)}
\end{eqnarray}
Note that the two fields corresponding to the diagram (ii),
$A^{\mbox{\scriptsize (ii-1)}}_{(\mu\nu),\rho}$
and 
$A^{\mbox{\scriptsize (ii-2)}}_{[\mu\nu],\rho}$,
are to satisfy the conditions  
\begin{equation}
A^{\mbox{\scriptsize (ii-1)}}_{(\mu\nu),\rho}+A^{\mbox{\scriptsize (ii-1)}}_{(\nu\rho),\mu}
+A^{\mbox{\scriptsize (ii-1)}}_{(\rho\mu),\nu} =0
\label{eq:n3Aii1cond}
\end{equation}
and
\begin{equation}
A^{\mbox{\scriptsize (ii-2)}}_{[\mu\nu],\rho}+A^{\mbox{\scriptsize (ii-2)}}_{[\nu\rho],\mu}
+A^{\mbox{\scriptsize (ii-2)}}_{[\rho\mu],\nu} =0 
\label{eq:n3Aii2cond}
\end{equation}
respectively.
Based on the above classification of the highest third rank tensor field $A_{\mu\nu\rho}$ for $n=3$ theory, the classification of the lower rank tensor fields $D_{\mu}^i$ ($i=1,2,3$) and the gauge parameter fields are determined.
The three vector fields part given by the second line of eq.(\ref{eq:phiADn3}) are recombined to the following form
\begin{eqnarray}
\tilde{\phi}_{n=3}(p)|_{D_{\mu}^i} &=&
\left(
\sum_{i=1}^3
\alpha_{(i)}^{\mu} (c_{(i+1)} b_{(i+2)} -  b_{(i+1)} c_{(i+2)}   )
\right)
 \ket{\downarrow\downarrow\downarrow  ;0,p} 
D^{\mbox{\scriptsize (i)}}_{\mu}(p) 
\\
&&\hspace*{-1cm}+
\left(
\alpha_{(1)}^{\mu} 
(c_{(2)} b_{(3)} \!-\!  b_{(2)} c_{(3)}   )
\!-\!
(c_{(1)} b_{(2)} \!-\!  b_{(1)} c_{(2)}   )
\alpha_{(3)}^{\mu} 
\right)
\ket{\downarrow \downarrow \downarrow ;0,p} 
D^{\mbox{\scriptsize (ii-1)}}_{\mu}(p) 
\\
&&\hspace*{-1cm}+
\Big(
\alpha_{(1)}^{\mu} 
(c_{(2)} b_{(3)} \!-\!  b_{(2)} c_{(3)}   )
\!+\!
(c_{(1)} b_{(2)} \!-\!  b_{(1)} c_{(2)}   )
\alpha_{(3)}^{\mu} 
-2 (c_{(1)} b_{(3)} \!-\!  b_{(3)} c_{(1)}   )
\alpha_{(2)}^{\mu} 
\Big)
\nonumber\\
&&\hspace*{5cm}
\times \ket{\downarrow \downarrow \downarrow ;0,p} 
D^{\mbox{\scriptsize (ii-2)}}_{\mu}(p) 
\end{eqnarray}
where the fields $D^{\mbox{\scriptsize (i)}}_{\mu}$, $D^{\mbox{\scriptsize (ii-1)}}_{\mu}$ and $D^{\mbox{\scriptsize (ii-2)}}_{\mu}$ respectively belong to the classes (i), (ii-1) and (ii-2).
Note that there is no $D_{\mu}$ field assigned for (iii).

Gauge parameter extended string field $\lambda_{n=3}$ for $n=3$ consists of three independent second rank tensor fields $\lambda_{\mu\nu}^i$ ($i=1,2,3$) and two scalar fields $\lambda_0^{j}$ ($j=1,2$).
They are classified by the four classes and the result is given as follows.
For the class (i), there is one type of symmetric second rank tensor field $\lambda^{\mbox{\scriptsize (i)}}_{(\mu\nu)}(p)$:
\begin{equation}
\tilde{\lambda}^{\mbox{\scriptsize (i)}}_{N=1}(p)
=
\sum_{i=1}^3
b_{(i)} \alpha_{(i+1)}^{(\mu} \alpha_{(i+2)}^{\nu)}  
 \ket{\downarrow\downarrow\downarrow  ;0,p} 
 \lambda^{\mbox{\scriptsize (i)}}_{(\mu\nu)}(p).
\end{equation}
For each of the classes (ii-1) and (ii-2), there are three type of gauge parameter fields: 
symmetric and anti-symmetric 2nd rank tensors, and a scalar.
Explicitly,  
\begin{eqnarray}
\tilde{\lambda}^{\mbox{\scriptsize (ii-1)}}_{N=1}(p)
&=&
\left(
\alpha_{(1)}^{(\mu} \alpha_{(2)}^{\nu)}b_{(3)}
-b_{(1)} \alpha_{(2)}^{(\mu}\alpha_{(3)}^{\nu)}
\right)
\ket{\downarrow \downarrow \downarrow ;0,p} 
i \lambda^{\mbox{\scriptsize (ii-1)}}_{(\mu\nu)}(p) 
\nonumber\\
& +&
\left(
\alpha_{(1)}^{[\mu} \alpha_{(2)}^{\nu]}b_{(3)}
+b_{(1)} \alpha_{(2)}^{[\mu}\alpha_{(3)}^{\nu]}
+ 2 b_{(2)} \alpha_{(1)}^{[\mu} \alpha_{(3)}^{\nu]} 
\right)
\ket{\downarrow \downarrow \downarrow ;0,p} 
i \lambda^{\mbox{\scriptsize (ii-1)}}_{[\mu\nu]}(p) 
\nonumber\\
& +&
\left(
b_{(1)} b_{(2)} c_{(3)} + c_{(1)} b_{(2)} b_{(3)} - 2 b_{(1)} c_{(2)} b_{(3)}
\right)
\ket{\downarrow \downarrow \downarrow ;0,p} 
i \lambda_0^{\mbox{\scriptsize (ii-1)}} (p) 
\end{eqnarray}
and
\begin{eqnarray}
\tilde{\lambda}^{\mbox{\scriptsize (ii-2)}}_{N=1}(p)
&=&
\left(
\alpha_{(1)}^{(\mu} \alpha_{(2)}^{\nu)}b_{(3)}
+b_{(1)} \alpha_{(2)}^{(\mu}\alpha_{(3)}^{\nu)}
-2 b_{(2)} \alpha_{(3)}^{(\mu}\alpha_{(1)}^{\nu)}
\right)
\ket{\downarrow \downarrow \downarrow ;0,p} 
i \lambda^{\mbox{\scriptsize (ii-2)}}_{(\mu\nu)}(p) 
\nonumber\\
& +&
\left(
\alpha_{(1)}^{[\mu} \alpha_{(2)}^{\nu]}b_{(3)}
-b_{(1)} \alpha_{(2)}^{[\mu}\alpha_{(3)}^{\nu]}
\right)
\ket{\downarrow \downarrow \downarrow ;0,p} 
i \lambda^{\mbox{\scriptsize (ii-2)}}_{[\mu\nu]}(p) 
\nonumber\\
& +&
\left(
b_{(1)} b_{(2)} c_{(3)} - c_{(1)} b_{(2)} b_{(3)} 
\right)
\ket{\downarrow \downarrow \downarrow ;0,p} 
i \lambda_0^{\mbox{\scriptsize (ii-2)}} (p) .
\end{eqnarray}
Finally, for the class (iii), 
\begin{equation}
\tilde{\lambda}^{\mbox{\scriptsize (iii)}}_{N=1}(p)
=
\sum_{i=1}^3
b_{(i)}\alpha_{(i+1)}^{[\mu} \alpha_{(i+2)}^{\nu]}  
 \ket{\downarrow\downarrow\downarrow  ;0,p} 
 \lambda^{\mbox{\scriptsize (iii)}}_{[\mu\nu]}(p).
\end{equation}

Now that we have prepared and classified all the fields and the gauge parameter fields for $n=3$, we are able to write the gauge transformation for each field and construct the gauge invariant action explicitly.
The explicit results will be given in the discussions of general $n$-rank tensor fields in the remainder of this section: 
The results for (i) and (iii) classes will be given in the next subsection~\ref{4.2} and the results for (ii-1) and (ii-2) classes will be given in subsection~\ref{4.3}.
Here, we only represent the explicit form of the Lagrangians with gauge transformations for the latter two classes (ii-1) and (ii-2), and give some comments on the relation between the two results which share the same Young diagram (ii).
The result is 
\begin{eqnarray}
{\cal L}^{\mbox{\scriptsize (ii-1)}}
&=&
\frac{3}{4} 
A^{\mbox{\scriptsize (1)}}_{(\mu\nu),\rho}
(\delta^{\rho}_{\sigma}\Box-  \partial^\rho \partial_{\sigma} ) 
A^{\mbox{\scriptsize (1)}}{}^{(\mu\nu),\sigma}
+ 
\frac{3}{2}
\partial_{\mu} A^{\mbox{\scriptsize (1)}}{}^{(\mu\nu),\rho}
\partial^{\sigma} A^{\mbox{\scriptsize (1)}}_{(\sigma\nu),\rho}
\nonumber\\
&& \quad 
-3 D^{\mbox{\scriptsize (1)}}_{\mu}
\partial_{\nu} \partial_{\rho} 
A^{\mbox{\scriptsize (1)}}{}^{(\mu\nu),\rho}
-2 
D^{\mbox{\scriptsize (1)}}_{\mu}
(\delta^{\mu\nu}\Box-  \partial^\mu \partial^{\nu} )
D^{\mbox{\scriptsize (1)}}_{\nu}
\label{eq:Lii1}
\end{eqnarray}
with
\begin{eqnarray}
\delta A^{\mbox{\scriptsize (1)}}_{(\mu\nu),\rho} &=&
\partial_{\rho}  \lambda^{\mbox{\scriptsize (1)}}_{(\mu\nu)} -
\partial_{(\rho}  \lambda^{\mbox{\scriptsize (1)}}_{\mu\nu)}
+ \partial_{\nu}\lambda^{\mbox{\scriptsize (1)}}_{[\mu\rho]}
+\partial_{\mu}\lambda^{\mbox{\scriptsize (1)}}_{[\nu\rho]},
\label{eq:trai11}
\\
\delta D^{\mbox{\scriptsize (1)}}_{\mu}
\label{eq:trai12}
&=&
\frac{1}{2} \partial^{\nu}  \lambda^{\mbox{\scriptsize (1)}}_{(\mu\nu)}
+\frac{3}{2} \partial^{\nu}  \lambda^{\mbox{\scriptsize (1)}}_{[\mu\nu]}
+\frac{3}{2} \partial_{\mu}  \lambda^{\mbox{\scriptsize (1)}}_{0},
\end{eqnarray}
and
\begin{eqnarray}
{\cal L}^{\mbox{\scriptsize (ii-2)}}
&=&
\frac{3}{4} 
A^{\mbox{\scriptsize (2)}}_{[\mu\nu],\rho}
(\delta^{\rho}_{\sigma}\Box-  \partial^\rho \partial_{\sigma} ) 
A^{\mbox{\scriptsize (2)}}{}^{[\mu\nu],\sigma}
+ 
\frac{3}{2}
\partial_{\mu} A^{\mbox{\scriptsize (2)}}_{[\mu\nu],\rho}
\partial^{\sigma} A^{\mbox{\scriptsize (2)}}_{[\sigma\nu],\rho}
\nonumber\\
&& \quad -6 D^{\mbox{\scriptsize (2)}}_{\mu}
\partial_{\nu} \partial_{\rho} 
A^{\mbox{\scriptsize (2)}}{}^{[\mu\nu],\rho}
-6 
D^{\mbox{\scriptsize (2)}}_{\mu}
(\delta^{\mu\nu}\Box-  \partial^\mu \partial^{\nu} )
D^{\mbox{\scriptsize (2)}}_{\nu}
\label{eq:Lii2}
\end{eqnarray}
with
\begin{eqnarray}
\delta A^{\mbox{\scriptsize (2)}}_{[\mu\nu],\rho} &=&
\partial_{\rho}  \lambda^{\mbox{\scriptsize (2)}}_{[\mu\nu]} -
\partial_{[\rho}  \lambda^{\mbox{\scriptsize (2)}}_{\mu\nu]}
+ \partial_{\mu}\lambda^{\mbox{\scriptsize (2)}}_{(\nu\rho)}
-\partial_{\nu}\lambda^{\mbox{\scriptsize (2)}}_{(\mu\rho)},
\\
\delta D^{\mbox{\scriptsize (2)}}_{\mu}
\label{eq:traii11}
&=&
\frac{1}{2} \partial^{\nu}  \lambda^{\mbox{\scriptsize (2)}}_{[\mu\nu]}
-\frac{1}{2} \partial^{\nu}  \lambda^{\mbox{\scriptsize (2)}}_{(\mu\nu)}
-\frac{1}{2} \partial_{\mu}  \lambda^{\mbox{\scriptsize (2)}}_{0}.
\label{eq:traii12}
\end{eqnarray}
We can show that 
these two Lagrangians are equivalent to each other if we identify the fields as
\begin{equation}
A^{\mbox{\scriptsize (2)}}_{[\mu\nu],\rho}
=\frac{1}{\sqrt{3}}\Big(
A^{\mbox{\scriptsize (1)}}_{(\rho\mu),\nu}
-
A^{\mbox{\scriptsize (1)}}_{[\rho\nu],\mu}
\Big)
,
\qquad
D^{\mbox{\scriptsize (2)}} = \frac{1}{\sqrt{3}}D^{\mbox{\scriptsize (1)}}
\end{equation}
and take into account the relations 
(\ref{eq:n3Aii1cond}) and (\ref{eq:n3Aii2cond}).
In general, gauge invariant actions for a particular Young diagram 
are equivalent to one another even if the appearance is different.
Thus we have completed the classification and the analysis of the $n=3$ theory.
For more general $n\ge 4$ cases, we can also classify the fields and gauge parameter fields according to the classification of Young diagrams of $n$ boxes and construct the corresponding gauge invariant actions.
Note that once the gauge invariant action is obtained, the gauge fixed action can also be constructed from the general one given in eq.(\ref{eq:Salpha}), though we do not go into detail about that.

\subsection{Totally symmetric and anti-symmetric tensor fields}
\label{4.2}
We extract the totally symmetric or anti-symmetric $n$-th rank tensor field part 
from the general $n$ theory and construct the corresponding gauge invariant action.
Each of these two cases corresponds to either of the two simplest Young diagrams with 
$n$ boxes of a single row or a single column.
For each case, we will collect all the fields
and the gauge parameter fields needed for constructing the gauge invariant action 
following the general procedure given previously.

For the symmetric tensor part, there is an 
$(n-2)$-th rank symmetric tensor field $D_{(\mu_1\cdots \mu_{n-2})}$
other than the $n$-th rank symmetric tensor field $A_{(\mu_1\mu_2\cdots \mu_n)}$.
These fields are collected in the form 
\begin{eqnarray}
\tilde{\phi}^{N=1}_{n, {\rm sym}}(p)
&=&
\alpha_{(1)}^{(\mu_1} \alpha_{(2)}^{\mu_2}\cdots \alpha_{(n)}^{\mu_n)}
\ket{\downarrow\downarrow \cdots \downarrow ;0,p} 
A_{ (\mu_1\mu_2\cdots \mu_n) } (p) 
\nonumber\\
& +&
\sum_{i<{j}}
\alpha_{(k_1)}^{(\mu_1} \alpha_{(k_2)}^{\mu_2}\cdots\alpha_{(k_{n-2})}^{\mu_{n-2}) }
(c_{(i)}b_{(j)}- b_{(i)}c_{(j)})
\ket{\downarrow\downarrow \cdots \downarrow ;0,p} 
\nonumber\\[-3pt]
&&\hspace*{8cm}\times D_{ (\mu_1\cdots \mu_{n-2}) }(p) 
\label{eq:nsymphi}
\end{eqnarray}
where the indices $ k_r$ ($r=1,\cdots, n-2$) is determined so as to satisfy $k_r< k_s$ for $r<s$ and $\{k_1 , \cdots,   k_{n-2}, i,j  \} =\{ 1,\cdots, n  \} $ for each $i$ and $j$ as for the general $\tilde{\phi}_n$ in eq.(\ref{eq:N1stategen}).
For the above $\tilde{\phi}^{N=1}_{n, {\rm sym}}$, we need only one type of gauge parameter field $\lambda_{ (\mu_1\mu_2\cdots \mu_{n-1}) }$ given by
\begin{equation}
\tilde{\lambda}^{N=1}_{n, {\rm sym}}(p)
= \frac{1}{n}
\sum_{j=1}^n
\alpha_{(k_1)}^{(\mu_1} \alpha_{(k_2)}^{\mu_2}\cdots\alpha_{(k_{n-1})}^{\mu_{n-1}) }
b_{(j)}
\ket{\downarrow\downarrow \cdots \downarrow ;0,p} 
i \lambda_{ (\mu_1\mu_2\cdots \mu_{n-1}) }(p) 
\end{equation}
where again $k_r< k_s$ for $r<s$ and $\{k_r, j \}=\{ 1,\cdots, n  \} $ for each $j$. 

Gauge transformation for the fields $A_{(\mu_1\cdots \mu_n)}$ and $D_{ (\mu_1\mu_2\cdots \mu_{n-2}) }$ is 
\begin{equation}
\delta A_{\mu_1\cdots \mu_n} = \partial_{(\mu_1} \lambda_{\mu_2\cdots \mu_{n})},
\qquad
\delta D_{\mu_1\cdots \mu_{n-2}} = \frac{1}{n} \partial^{\mu} 
\lambda_{\mu \mu_1 \cdots \mu_{n-2}}.
\end{equation}
The Lagrangian is obtained by substituting eq.(\ref{eq:nsymphi}) into 
eq.(\ref{eq:N1action}) and calculating the inner products explicitly.
The result is 
\begin{eqnarray}
{\cal L}^{n, {\rm sym}}
&=&
-\frac{1}{2} \partial_\nu A_{\mu_1\cdots \mu_n}  \partial^\nu A^{\mu_1\cdots \mu_n} 
+n(n-1)  \partial_\nu D_{\mu_1\cdots \mu_{n-2}}  \partial^\nu D^{\mu_1\cdots \mu_{n-2}}  
\nonumber\\
&&+\frac{n}{2} \partial_\mu A^{\mu \mu_1\cdots \mu_{n-1}} 
 \partial^\nu A_{\nu \mu_1\cdots \mu_{n-1}}  
+n(n-1) D_{\mu_1\cdots \mu_{n-2}} \partial_\mu \partial_\nu   
A^{\mu\nu\mu_1\cdots \mu_{n-2}}
\nonumber\\
&& + \frac{n(n-1)(n-2)}{2} \partial_\mu D^{\mu \mu_1\cdots \mu_{n-3}} 
 \partial^\nu D_{\nu \mu_1\cdots \mu_{n-3}}  
.
\end{eqnarray}
As we have expected, this Lagrangian exactly reproduces the so called `triplet' action 
which is obtained by taking the tensionless ($\alpha'\rightarrow \infty$) limit of the open bosonic string field theory~\cite{Asano:2012qn,Bengtsson:1986ys,Ouvry:1986dv,Francia:2002pt, Sagnotti:2003qa}.
Note that this action contains not a triplet of fields, but only a doublet of $A$ and $D$. 
This is because we start from the minimal action $S_n^{\rm min}$ which does not contain auxiliary extended string field $b_0 \Phi_n$ from the beginning.
The gauge fixed action can also be constructed from the general one given in eq.(\ref{eq:SFPgf}) and the result is the same as that given in ref.\cite{Asano:2012qn}.

\bigskip

Next, we consider the case of $n$-th rank anti-symmetric tensor part.
In this case, there is no extra lower rank field involved in the original $A_{ [\mu_1\mu_2\cdots \mu_n] } $ field.
Thus, the field in the form of the extended string field is 
\begin{equation}
\tilde{\phi}^{N=1}_{n, {\rm anti-sym}}(p)
=
\alpha_{(1)}^{[\mu_1} \alpha_{(2)}^{\mu_2}\cdots \alpha_{(n)}^{\mu_n]}
\ket{\downarrow\downarrow \cdots \downarrow ;0,p} 
A_{ [\mu_1\mu_2\cdots \mu_n] } (p) ,
\end{equation}
and the gauge parameter field has the form 
\begin{equation}
\tilde{\lambda}^{N=1}_{n, {\rm anti-sym}}(p)
= \frac{1}{n}
\sum_{j=1}^n
(-)^{j-1}
\alpha_{(k_1)}^{(\mu_1} \alpha_{(k_2)}^{\mu_2}\cdots\alpha_{(k_{n-1})}^{\mu_{n-1}) }
b_{(j)}
\ket{\downarrow\downarrow \cdots \downarrow ;0,p} 
i \lambda_{ (\mu_1\mu_2\cdots \mu_{n-1}) }(p) 
\end{equation}
where $k_r<k_s$ for $r<s$ and $\{k_r, j \}=\{ 1,\cdots, n  \} $ for each $j$. 
Gauge transformation is given by
\begin{equation}
\delta A_{[\mu_1\cdots \mu_n]} = \partial_{[\mu_1} \lambda_{\mu_2\cdots \mu_{n}]}.
\label{eq:nantisymgt}
\end{equation}
The Lagrangian for the field $A_{ [\mu_1\mu_2\cdots \mu_n] } $ is then written as
\begin{equation}
{\cal L}^{n, {\rm anti-sym}}
=
-\frac{1}{4(n+1)} H_{\mu_1\mu_2\cdots\mu_{n+1}}H^{\mu_1\mu_2\cdots\mu_{n+1}  }
\label{eq:antisymL}
\end{equation}
where 
\begin{equation}
H_{\mu_1\mu_2\cdots\mu_{n+1}} =(n+1) \partial_{[\mu_1}A_{ \mu_2\cdots\mu_{n+1]} }.
\end{equation}
Gauge fixed action for eq.(\ref{eq:antisymL}) can also be constructed from eq.(\ref{eq:SFPgf}). 
To fix the gauge symmetry (\ref{eq:nantisymgt}) completely, 
we need up to $n$-th order of (anti-)ghost fields with the Lagrange multiplier fields in the gauge fixed action.
Explicitly, $m$-th order (anti-)ghost fields ($m=1,\cdots,n$)
are provided by
the SU(1,1)-spin$=\frac{m}{2}$ part of $\phi_n^{(-m+2k)}$ ($k=0,1,\cdots,m $) 
with the Lagrange multiplier fields $\beta_n^{(m-2l)}$ ($l=0,1,\cdots, m-1 $).
Note that in this case all the fields belonging to the SU(1,1)-spin$=\frac{m}{2}$ part are  $(n-m)$-th rank anti-symmetric tensor fields. 
These structures naturally coincide with the known results of the gauge fixing problem for anti-symmetric tensor field theory given in the literature~\cite{Kimura:1980zd_p3}. 

Finally, note that the gauge invariant and the gauge fixed actions for the anti-symmetric tensor field given above can also be deduced from the tensionless limit of the $n=1$ minimal open string field theory action as is shown in~ref.\cite{Asano:2012qn}.
However, as we have seen, to obtain the same action,
it is more straightforward to use our extended string field theory than to use the open string field theory.
For more general mixed symmetric case, we see that it is a great advantage using our extended string field theory to obtain the consistent gauge invariant and gauge fixed actions since it is very complicated to extract the same actions from the tensionless limit of the open string field theory.

\subsection{Simple mixed symmetric fields}
\label{4.3}
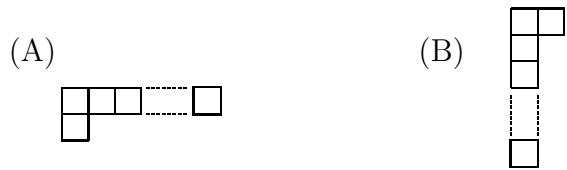
\begin{figure}[ht]
  \centering
\begin{picture}(200,70)(0,0)
\put(-10,10){\line(0,1){20}}
\put(-10,20){\line(1,0){30}}
\put(-10,30){\line(1,0){30}}
\put(20,20){\line(0,2){10}}
\put(0,10){\line(0,1){20}}
\put(-10,10){\line(1,0){10}}
\put(10,20){\line(0,1){10}}
\multiput(22,20)(2,0){8}{\line(1,0){1}}
\multiput(22,30)(2,0){8}{\line(1,0){1}}
\put(40,20){\framebox(10,10){}}
\put(-30,40){{(A)}}
\put(125,40){{(B)}}
\put(160,30){\line(0,1){30}}
\put(170,30){\line(0,1){30}}
\put(160,30){\line(1,0){10}}
\put(160,50){\line(1,0){20}}
\put(160,60){\line(1,0){20}}
\put(180,50){\line(0,1){10}}
\put(160,40){\line(1,0){10}}
\put(160,0){\framebox(10,10){}}
\multiput(160,12)(0,2){8}{\line(0,1){1}}
\multiput(170,12)(0,2){8}{\line(0,1){1}}
\end{picture}
  \caption{Young diagrams for $n$-th rank mixed symmetric tensor fields (A)~$A_{(\mu_1\cdots \mu_{n-1}),\mu_n }$ and (B)~$B_{[\mu_1\cdots \mu_{n-1}],\mu_n }$}
  \label{fig:mixedsymAB}
\end{figure}

We discuss the example of the two simplest $n$-th rank mixed symmetric tensor fields identified by the Young diagrams depicted in fig.\ref{fig:mixedsymAB}.

\paragraph{The action for the diagram (A)}
The $n$-th rank tensor field corresponding to the Young diagram (A) can be chosen to have the form $A_{(\mu_1\cdots \mu_{n-1}),\mu_n }$ where only the first $n-1$ indices are symmetric and the condition 
\begin{equation}
\sum_{i=1}^n A_{(\mu_1\cdots \hat{\mu_i} \cdots\mu_{n}),\mu_i } 
\Big(=  n A_{((\mu_1\cdots \mu_{n-1}),\mu_n )}\Big) =0 
\label{eq:mixAAcond}
\end{equation}
is satisfied.
Note that we can choose another form of tensor field for representing the same diagram.
We chose this particular representation of the field given in eq.(\ref{eq:mixAAcond}) since it is simple and easy for analyzing the action. 
As we have seen in the example of $n=3$ case, we obtain the equivalent action if we start from the different representation of the field as long as we deal with the same Young diagram.
In order to extract the gauge invariant action including the field $A_{(\mu_1\cdots \mu_{n-1}),\mu_n }$ from the general action,
we have to incorporate additional two types of $(n\!-\!2)$-th rank tensor fields 
$D^{\rm S}_{(\mu_1\cdots \mu_{n-3}),\mu_{n-2} }$ and
$D^{\rm M}_{(\mu_1\cdots \mu_{n-2})}$.
Here $D^{\rm S}$ is a totally symmetric tensor field and 
$D^{\rm M}$ is a mixed-symmetric tensor field which must satisfy the condition similar to eq.(\ref{eq:mixAAcond}):
$D^{\rm M}_{((\mu_1\cdots \mu_{n-3}),\mu_{n-2} )}=0$.
These two fields with $A_{(\mu_1\cdots \mu_{n-1}),\mu_n }$ are collected in the following extended string field as
\begin{eqnarray}
&& \hspace{-7mm}\tilde{\phi}^{N=1}_{(n-1)+1}(p)
\nonumber\\
&&=
\left\{
\alpha_{(1)}^{(\mu_1}\cdots \alpha_{(n-1)}^{\mu_{n-1})} \alpha_{(n)}^{\mu}
-
\alpha_{(1)}^{\mu}\alpha_{(2)}^{(\mu_1}\cdots \alpha_{(n)}^{\mu_{n-1})} 
\right\}
\ket{\downarrow\downarrow \cdots \downarrow ;0,p} \,
A_{ (\mu_1\mu_2\cdots \mu_{n-1}),\mu } (p) 
\nonumber\\
&&
+
\sum_{i=2}^{n-1}
\Big\{
\alpha_{(k_1)}^{(\mu_1}\cdots \alpha_{(k_{n-2})}^{\mu_{n-2)}}
(c_{(i)}b_{(n)}\!-\! b_{(i)}c_{(n)})
\nonumber\\
&&\qquad\qquad
 -
\alpha_{(k'_1)}^{(\mu_1}\cdots \alpha_{(k'_{n-2})}^{\mu_{n-2})} 
(c_{(1)}b_{(i)}\!-\! b_{(1)}c_{(i)})
\Big\}
\,\ket{\downarrow\downarrow \cdots \downarrow ;0,p} \,
D^{\rm S}_{ (\mu_1\cdots \mu_{n-2}) }(p) 
\nonumber\\
& &+
\Bigg(
\sum_{1\le i<{j}<n}
\alpha_{(l_1)}^{(\mu_1} \cdots\alpha_{(l_{n-3})}^{\mu_{n-3}) }
(c_{(i)}b_{(j)}- b_{(i)}c_{(j)})\alpha^{\nu}_{(n)}
\nonumber\\
&&\qquad -
\sum_{1< i<{j}\le n}
\alpha^{\nu}_{(1)}\alpha_{(l'_1)}^{(\mu_1} \cdots\alpha_{(l'_{n-3})}^{\mu_{n-3}) }
(c_{(i)}b_{(j)}- b_{(i)}c_{(j)})
\nonumber\\
&&\qquad +
\sum_{i=2}^{n-1}
\Big\{
\alpha_{(k_1)}^{(\mu_1}\cdots \alpha_{(k_{n-3})}^{\mu_{n-3}}
\alpha_{(k_{n-2})}^{\nu)} 
(c_{(i)}b_{(n)}\!-\! b_{(i)}c_{(n)})
\nonumber\\
&&
\qquad\quad -
\alpha_{(k'_1)}^{(\mu_1}\cdots \alpha_{(k'_{n-3})}^{\mu_{n-3}} 
\alpha_{(k'_{n-2})}^{\nu)} 
(c_{(1)}b_{(i)}\!-\! b_{(1)}c_{(i)})
\Big\}\Bigg)
\,\ket{\downarrow\downarrow \cdots \downarrow ;0,p} \,
D^{\rm M}_{ (\mu_1\cdots \mu_{n-3}),\nu }(p) 
\label{eq:phimixedsym}
\end{eqnarray}
where again indices $k$, $k'$, $l$ and $l'$ are determined in order that  
each set $\{l_1 , \cdots,   l_{n-3}, i,j,n  \}$,
$\{1, l'_1 , \cdots,   l'_{n-3}, i,j  \}$,
$\{k_1 , \cdots,   k_{n-2}, i,n  \}$, or
$\{1, k'_1 , \cdots,   k'_{n-2}, i \}$ coincides with 
the set of $n$ integers $\{ 1,\cdots, n  \} $ for any $i$ or $(i,j)$.
Also, $k_r<k_s$ for any $r<s$, and the similar relations hold for indices $k'$, $l$ and $l'$.

For these fields, three types of gauge parameter fields are involved:
two types of $(n\!-\!1)$-th rank tensor fields 
$\lambda^{\rm S}_{(\mu_1\cdots\mu_{n-1})}$ and $\lambda^{\rm M}_{(\mu_1\cdots\mu_{n-2}),\mu_{n-1}}$,
and a symmetric $(n\!-\!3)$-th rank tensor field
$\tilde{\lambda}^{\rm S}_{(\mu_1\cdots\mu_{n-3})}$.
Gauge transformation for each field in the coordinate representation becomes 
\begin{eqnarray}
\delta A_{ (\mu_1\cdots \mu_{n-1}),\nu }&=&
\partial_{\nu}\lambda^{\rm S}_{(\mu_1\cdots\mu_{n-1})}
-\partial_{(\nu}\lambda^{\rm S}_{\mu_1\cdots\mu_{n-1})}
+(n-1)\partial_{(\mu_{n-1}}\lambda^{\rm M}_{\mu_1\cdots\mu_{n-2}),\nu},
\\
\delta  D^{\rm M}_{ (\mu_1\cdots \mu_{n-3}),\nu }&=&
\partial^{\rho}\lambda^{\rm M}_{\rho\mu_1\cdots\mu_{n-3},\nu}
-\partial^{\rho}\lambda^{\rm M}_{\rho(\mu_1\cdots\mu_{n-3},\nu)}
- \partial_{\nu}\tilde{\lambda}^{\rm S}_{(\mu_1\cdots\mu_{n-3})}
+\partial_{(\nu}\tilde{\lambda}^{\rm S}_{\mu_1\cdots\mu_{n-3})},
\\
\delta D^{\rm S}_{ \mu_1\cdots \mu_{n-2} }&=&
\frac{1}{2}\partial^{\nu} \lambda^{\rm S}_{(\mu_1\cdots\mu_{n-2}\nu)}
+\frac{n}{2(n-2)} \partial^{\nu} \lambda^{\rm M}_{(\mu_1\cdots\mu_{n-2}),\nu}
+\frac{n}{2}\partial_{(\mu_{n-2}} \tilde{\lambda}^{\rm S}_{\mu_1\cdots\mu_{n-3})}.
\end{eqnarray}
The gauge invariant action can be calculated by substituting eq.(\ref{eq:phimixedsym}) into eq.(\ref{eq:N1action}).
After simplifying the result by performing some suitable partial integrations, 
the Lagrangian becomes
\begin{eqnarray}
{\cal L}^{(n-1),1}&\!=\!&
\frac{n}{2(n\!-\!1)} A_{\mu_1\cdots\mu_{n-1},\nu }
( \Box \delta^{\nu}{}_{\rho}-  \partial_\rho \partial^{\nu} ) 
A^{\mu_1\cdots\mu_{n-1},\rho }
+ \frac{n}{2}\partial^{\rho} A_{\mu_1\cdots\mu_{n-2}\rho,\sigma }
\partial_{\nu} A^{\mu_1\cdots\mu_{n-2}\nu,\sigma }
\nonumber\\
&&\hspace*{-1.5cm}+n(n-2)
(D^{\rm S}_{\mu_1\cdots\mu_{n-3}\nu} - D^{\rm M}_{\mu_1\cdots\mu_{n-3},\nu})
 \partial_\rho\partial_\sigma A^{\rho\sigma\mu_1\cdots\mu_{n-3},\nu}
\nonumber\\
&&\hspace*{-1.5cm}-2(n-2) D^{\rm S}_{\mu_1\cdots\mu_{n-3}\nu }
( \Box \delta^{\nu}{}_{\rho}-  \partial_\rho \partial^{\nu} ) 
D^{\rm S\,}{}^{\mu_1\cdots\mu_{n-3}\rho }
+2n(n-2) \partial^{\rho} D^{\rm M}_{\mu_1\cdots\mu_{n-3},\rho }
\partial_{\nu} D^{\rm S\,}{}^{\mu_1\cdots\mu_{n-3}\nu }
\nonumber\\
&&\hspace*{-1.5cm}-\frac{1}{2}n(n-2) D^{\rm M}_{\mu_1\cdots\mu_{n-3},\nu } 
( 2\Box \delta^{\nu}{}_{\rho}-  \partial_\rho \partial^{\nu} ) 
D^{\rm M\,}{}^{\mu_1\cdots\mu_{n-3},\rho }
\nonumber\\
&&\hspace*{-1.5cm}+\frac{1}{2}n(n-2)(n-3)\partial^\nu D^{\rm M}_{\nu\mu_1\cdots\mu_{n-4},\mu_{n-3} }
\partial_\rho D^{\rm M\,}{}^{\rho\mu_1\cdots\mu_{n-4},\mu_{n-3} }.
\end{eqnarray}
Note that for $n=3$, this Lagrangian 
is reduced to the one given in eq.(\ref{eq:Lii1}).

\paragraph{The action for the diagram (B)}
The $n$-th rank tensor field corresponding to the diagram (B) can be taken as 
the form $B_{[\mu_1\cdots \mu_{n-1}],\mu_n }$ which 
satisfies the condition
\begin{equation}
\sum_{i=1}^n(-)^{n-i} B_{[\mu_1\cdots \hat{\mu}_i\cdots \mu_n],\mu_i}
\Big(= n B_{[[\mu_1\cdots \mu_{n-1}],\mu_n ]} \Big)=0.
\end{equation}
To construct the gauge invariant action for this field from the general action 
eq.(\ref{eq:N1action}), we have to include a totally anti-symmetric $(n\!-\!2)$-th rank tensor field $E_{ [\mu_1\mu_2\cdots \mu_{n-2}] }$.  
These two fields are collected in the form of the extended string field as 
\begin{eqnarray}
&& \!\!\!\tilde{\phi}^{N=1}_{[n-1]+1}(p)
\nonumber\\
&&=
\left(
\alpha_{(1)}^{[\mu_1}\cdots \alpha_{(n-1)}^{\mu_{n-1}]} \alpha_{(n)}^{\mu}
+(-)^n 
\alpha_{(1)}^{\mu}\alpha_{(2)}^{[\mu_1}\cdots \alpha_{(n)}^{\mu_{n-1}]} 
\right)
\ket{\downarrow\downarrow \cdots \downarrow ;0,p} 
B_{ [\mu_1\mu_2\cdots \mu_{n-1}],\mu } (p) 
\nonumber\\
& &+
\!\Bigg(
\sum_{i=2}^{n-1}
\Big\{
(-)^{n\!-\!1\!-\!i}
\alpha_{(k_1)}^{[\mu_1}\cdots \alpha_{(k_{n-2})}^{\mu_{n-2}]} 
(c_{(i)}b_{(n)}\!-\! b_{(i)}c_{(n)})
+ 
(-)^{i}
\alpha_{(k'_1)}^{[\mu_1}\cdots \alpha_{(k'_{n-2})}^{\mu_{n-2}]} 
(c_{(1)}b_{(i)}\!-\! b_{(1)}c_{(i)})
\Big\}
\nonumber\\
&&\qquad+ 
2 (-)^{n}\alpha_{(2)}^{[\mu_1}  \cdots 
\alpha_{(n-1)}^{\mu_{n-2}]} 
(c_{(1)}b_{(n)}- b_{(1)}c_{(n)})
\Bigg)
\ket{\downarrow\downarrow \cdots \downarrow ;0,p} 
E_{ [\mu_1\mu_2\cdots \mu_{n-2}] }(p) 
\end{eqnarray}
where $k_r$ or $k'_r$ are uniquely chosen for each $i$ so that 
$\{k_1 , \cdots,   k_{n-2}, i,n  \}$ or
$\{1, k'_1 , \cdots,   k'_{n-2}, i \}$ coincides with 
the set of $n$ integers $\{ 1,\cdots, n  \} $ with the condition 
$k_r<k_s$ or $k'_r<k'_s$ for $r<s$.
 
For these fields, the following three types of gauge parameter fields are involved in the gauge transformation:
two types of $(n\!-\!1)$-th rank tensor fields 
$\lambda^{\rm A}_{[\mu_1\cdots\mu_{n-1}]}$ and 
$\lambda^{\rm {M}'}_{[\mu_1\cdots\mu_{n-2}],\mu_{n-1}}$,
and an anti-symmetric $(n\!-\!3)$-th rank tensor 
$\tilde{\lambda}^{\rm A}_{[\mu_1\cdots\mu_{n-3}]}$.
Here the field $\lambda^{\rm {M}'}$ satisfies the condition
\begin{equation}
\sum_{i=1}^{n-1}(-)^{n-1-i} 
\lambda^{\rm {M}'}_{[\mu_1\cdots\hat{\mu_i} \cdots\mu_{n-2}],\mu_{i}}
\,\Big(=
(n-1)
\lambda^{\rm {M}'}_{[[\mu_1\cdots\mu_{n-2}],\mu_{n-1}]}
\Big)
=0.
\label{eq:lambdan-11B}
\end{equation}
Then the gauge transformation for each field in the coordinate representation is
given by
\begin{eqnarray}
\delta B_{ [\mu_1\cdots \mu_{n-1}],\nu }&=&
\partial_{\nu}\lambda^{\rm A}_{[\mu_1\cdots\mu_{n-1}]}
-\partial_{[\nu}\lambda^{\rm A}_{\mu_1\cdots\mu_{n-1}]}
+(n-1)\partial_{[\mu_{1}}\lambda^{\rm {M}'}_{\mu_2\cdots\mu_{n-1}],\nu}\;,
\\
\delta E_{ [\mu_1\cdots \mu_{n-2}] }&=&
\frac{1}{2}\partial^{\nu} \lambda^{\rm A}_{[\mu_1\cdots\mu_{n-2}\nu]}
+\frac{1}{2} (-)^n \partial^{\nu} \lambda^{\rm {M}'}_{[\mu_1\cdots\mu_{n-2}],\nu}
+\frac{3}{2}(n-2) \partial_{[\mu_{n-2}} \tilde{\lambda}^{\rm A}_{\mu_1\cdots\mu_{n-3}]}\;.
\end{eqnarray}
Then the Lagrangian for the field $B_{[\mu_1\cdots\mu_{n-1}],\nu }$ with 
$E^{[\mu_1\cdots\mu_{n-2}]}$ becomes after performing some suitable partial integrations
\begin{eqnarray}
{\cal L}^{[n-1],1}
&=&
\frac{n}{2}
\Bigg(
\frac{1}{n-1} B_{[\mu_1\cdots\mu_{n-1}],\nu }
( \Box \delta^{\nu}{}_{\rho}-  \partial_\rho \partial^{\nu} ) 
B^{[\mu_1\cdots\mu_{n-1}],\rho }
\nonumber\\
&& + 
\partial^{\rho} B_{[\mu_1\cdots\mu_{n-2}\rho],\sigma }
\partial_{\nu} B^{[\mu_1\cdots\mu_{n-2}\nu],\sigma }
+ 4  E_{[\mu_1\cdots\mu_{n-2}]}
\partial_{\nu} \partial_{\rho} B^{[\mu_1\cdots\mu_{n-2}\nu],\rho }
\nonumber\\
&&
- 4 E_{[\mu_1\cdots\mu_{n-2}]}  \Box 
E_{[\mu_1\cdots\mu_{n-2}]}
-4(n-2) 
\partial^{\rho} E_{[\rho \mu_1\cdots\mu_{n-3}]}
\partial_{\nu} E^{[\nu\mu_1\cdots\mu_{n-3}]}
\Bigg).
\label{eq:LmixedB}
\end{eqnarray}
Note that this Lagrangian is only nontrivial for $D>n$ as a mixed symmetric field theory since the $B$ field does not exist for $D\le n-2$, and it is reduced to traceless second rank tensor field or vector field for $D= n$ or $D=n-1$.
Note also that eq.(\ref{eq:LmixedB}) is reduced to the one given in eq.(\ref{eq:Lii2}) for $n=3$.

We can show that the above Lagrangian given in eq.(\ref{eq:LmixedB}) is divided into two independent parts if we use the two fields $B'_{[\mu_1\cdots\mu_{n-1}],\nu} $
and $E'_{[\mu_1\cdots\mu_{n-2}]}$ defined by 
\begin{eqnarray}
E'_{[\mu_1\cdots\mu_{n-2}]} &=&
E_{[\mu_1\cdots\mu_{n-2}]} -\frac{1}{2} B_{[\mu_1\cdots\mu_{n-2}\rho],\nu }
\eta^{\rho\nu},
\\
B'_{[\mu_1\cdots\mu_{n-1}],\nu} &=&
B_{[\mu_1\cdots\mu_{n-1}],\nu} 
+ \frac{2(n-1)}{D-n} 
E'_{[\mu_1\cdots\mu_{n-2}} \eta_{\mu_{n-1}] \rho}
\end{eqnarray}
instead of the original $B$ and $E$ fields.
The resulting Lagrangian can be written as
\begin{equation}
 {\cal L}^{[n-1],1}= {\cal L}^{[n-1],1}|_{B'} + {\cal L}^{[n-1],1}|_{E'}
\end{equation}
where the $B'$ part becomes 
\begin{eqnarray}
{\cal L}^{[n-1],1}|_{B'} 
&=&
\frac{n}{2}
\Bigg(
\frac{1}{n-1} B'_{[\mu_1\cdots\mu_{n-1}],\nu }
( \Box \delta_{\rho}{}^{\nu} -  \partial_\rho \partial^{\nu} ) 
B^{'[\mu_1\cdots\mu_{n-1}],\rho }
\nonumber\\
&& + 
\partial^{\rho} B'_{[\mu_1\cdots\mu_{n-2}\rho],\sigma }
\partial_{\nu} B^{'[\mu_1\cdots\mu_{n-2}\nu],\sigma }
-  B'_{[\mu_1\cdots\mu_{n-2}\sigma],}{}^{\sigma }
 (\Box \delta_{\rho\nu} -  2\partial_\rho \partial_{\nu})
 B^{'[\mu_1\cdots\mu_{n-2}\rho],\nu}
\nonumber\\
&&  
- (n-2)  \partial^\nu  B'_{[\nu\mu_1\cdots\mu_{n-3}\sigma],}{}^{\sigma }
\partial_\rho  B^{'[\rho\mu_1\cdots\mu_{n-3}\gamma],}{}_{\gamma }
\Bigg),
\label{eq:LmixedAn-11}
\end{eqnarray}
while the totally antisymmetric tensor $E'$ part becomes 
\begin{equation}
{\cal L}^{[n-1],1}|_{E'}
=
-\frac{2n}{(D-n)(n-1)} H'_{\mu_1\mu_2\cdots\mu_{n-1}}H'{}^{\mu_1\mu_2\cdots\mu_{n-1}  }
\end{equation}
with
\begin{equation}
H'_{\mu_1\mu_2\cdots\mu_{n-1}}=(n-1)\partial_{[\mu_1}E'_{\mu_2\cdots\mu_{n-1}]}.
\end{equation}
The former Lagrangian ${\cal L}^{[n-1],1}|_{B'} $ is invariant under the gauge transformation 
\begin{equation}
\delta B'_{[\mu_1\cdots\mu_{n-1}],\nu } 
=\partial_\nu \lambda'{}^{\rm A}_{ \mu_1\cdots\mu_{n-1} }
-\partial_{[\nu}\lambda'{}^{\rm A}_{\mu_1\cdots\mu_{n-1}] }
 +\partial_{[\mu_1}\lambda'{}^{\rm M}_{\mu_2\cdots\mu_{n-1}] ,\nu}\;.
\end{equation}
Here, there are two independent 
gauge parameter fields given by $(n-1)$-th rank tensor fields:
$\lambda'{}^{\rm A}_{\mu_1\cdots\mu_{n-1}  }$ and $\lambda'{}^{\rm M}_{[\mu_1\cdots\mu_{n-2}],\mu_{n-1}  }$. 
The latter mixed-symmetric field satisfies the same relation as for $\lambda^{\rm M}_{[\mu_1\cdots\mu_{n-2}],\mu_{n-1}  }$ given in eq.(\ref{eq:lambdan-11B}).
Note that the Lagrangian (\ref{eq:LmixedAn-11}) is equivalent to the one given in the literature~\cite{Aulakh:1986cb}. 
In the literature, the structure of the gauge fixed action in terms of BRST transformation has been also discussed. 
Our gauge fixed action according to eq.(\ref{eq:SFPgf}) naturally reproduces the same result: For the $B_{[\mu_1\cdots\mu_{n-1}],\nu }$ (or $B'_{[\mu_1\cdots\mu_{n-1}],\nu }$) field, we see that there appear up to $(n-1)$-th order (anti-)ghost fields $\phi^{(l)}_n$ with appropriate number of Lagrange multiplier fields $\beta^{(l')}_n$ in the gauge fixed action reduced from eq.(\ref{eq:SFPgf}).
On the other hand, the latter Lagrangian ${\cal L}^{[n-1],1}|_{E'} $ 
is equivalent to eq.(\ref{eq:antisymL}) for $D>n$. 

\subsection{General mixed symmetric fields}
We briefly comment on the structure of gauge invariant and gauge fixed actions for general mixed symmetric fields. 

Consider a Young diagram of $n$ boxes with $k$ rows and the length of $i$-th row $m_i$.
Note that the relations $n=\sum_{i=1}^k m_i$ and $m_i \ge m_j$ for $ i < j$ are  satisfied.
This diagram is often represented by $(m_1, m_2,\cdots, m_k)$. 
To construct the gauge invariant action for the diagram, 
we first choose a basis state $\ket{\mu_1,\cdots,\mu_n;p}$ of the form given in eq.(\ref{eq:basisstn}) for the highest rank $n$ tensor field $A_{\mu_1,\cdots,\mu_n}^{(m_1, m_2,\cdots, m_k)}(p)$.
Then, the basis states of the rank $n-2$ tensor fields which should be included in the gauge invariant action for the field $A_{\mu_1,\cdots,\mu_n}^{(m_1, m_2,\cdots, m_k)}$ are constructed roughly by replacing every possible pair of symmetric oscillators $\alpha_{(i)}$ and $\alpha_{(j)}$ in $\ket{\mu_1,\cdots,\mu_n;p}$ with $ b_{(i)}c_{(j)} - c_{(i)}b_{(j)} $ and summing up the result. Furthermore, the basis states for further lower rank $n-4$, $n-6$, $\cdots$ tensor fields are constructed by repeating the same processes.  
Note that in the above set of processes, we cannot replace four (or more) totally symmetric $\alpha$ oscillators with two $b$'s and $c$'s in such a way since the result vanishes identically.
This can be seen by the relation 
\begin{equation}
\sum_{{\rm perms.\, of\,}{\{ijkl\}}} ( b_{(i)}c_{(j)} - c_{(i)}b_{(j)} )(b_{(k)}c_{(l)} - c_{(k)}b_{(l)} ) = 0.
\end{equation}
Thus, in general, the rank of the lowest rank tensor fields included in the gauge fixed action is determined as $n-2l$ where $l$ is the number of rows of the relevant Young diagram whose length is greater than or equal to 2. 
More precisely, for the diagram $(m_1, m_2,\cdots, m_k)$, $l$ is determined by the condition $m_l\ge 2$ and $m_{l+1}\le 1$.

As for the gauge fixed action, the chain of (anti-)ghost fields are determined by the number of rows of the Young diagram as we have mentioned in the case of anti-symmetric tensor field $B_{[\mu_1\cdots \mu_n]}$ or the mixed symmetric field $B_{[\mu_1\cdots \mu_{n-1}],\mu_n}$.
Explicitly, the gauge fixed action for the diagram with $k$ rows should contain up to $k$-th order (anti-)ghost fields.

\section{Discussions}

We have constructed both gauge invariant and gauge fixed free actions for extended string field theory based on extended state space which is made of $n$-copies of open string state space. The field theory possesses higher spin gauge fields in its massless spectrum.


Note that as we have seen in the previous section,
compared to the tensionless limit of the open string theory, our extended theory enables us to derive the action for massless tensor fields of any symmetry in a relatively simple way. The resultant action has desirable properties such as (a) covariant, local and needs no extra constraints for the fields, (b) the physical spectrum is apparent and the no-ghost theorem is satisfied, (c) the gauge fixed action corresponding to various covariant gauges is derived automatically, and so on.
The so-called `triplet' action, or doublet action in our representation, has been almost unique concrete example for massless higher spin fields derived systematically from the tensionless limit of the open bosonic string field theory.
This triplet action has inspired various analyses of higher spin field theory because of its simple but prominent structure~\cite{Francia:2002pt, Sagnotti:2003qa, Francia:2010qp, Fotopoulos:2009iw, Buchbinder:2007ak, Sorokin:2008tf}.
We thus naturally expect that the action for more general massless mixed symmetric fields  gives additional outcome for the study of higher spin theory.  
For such a purpose, it is much easier to use our setup of extended string field theory. 
It would be interesting to proceed further to such direction and to discuss the relation with preceding works~\cite{Sagnotti:2003qa,Chung:1987mv,Labastida:1986gy,Campoleoni:2008jq,Campoleoni:2009je,Campoleoni:2012th,Buchbinder:2011xw}.

Furthermore, though we have only dealt with the massless part of the action in detail, it is possible to discuss the structure of the gauge invariant and gauge fixed actions for general massive tensor fields within the framework of the extended string field theory if we fix $D=26$.
Structure of massive higher spin field theory is more complicated than the massless one since we in general have to include the St\"{u}ckelberg-like fields to support the gauge invariance for the original fields.
We may be able to well understand the structure of such St\"{u}ckelberg-like fields in general by utilizing our theory where the action for general massive fields are simpler than the one appearing in open string field theory if we choose appropriate $n$.

We may also be able to extend the superstring field theory by assembling multiple copies of the theory at least for quadratic part of the action.
If such an extension is accomplished, we expect that we can consistently construct the appropriate actions for general fermionic fields with half-integer spin. 

Now our next step is to construct gauge invariant interaction terms. A natural guess for their form is simply a tensor-product of the open string cubic vertex and necessary modifications if any. For $n=2$ case, free spectrum coincides with that of closed string, therefore if we properly follow the geometric picture of closed string world sheet, then we will eventually obtain non-polynomial interactions of closed string field theory so as to cover its moduli space~\cite{Saadi:1989tb,Kugo:1989aa}. However, as for the other $n$, we do not have such an intuition on the moduli space of the amplitude a priori. Even for $n=2$ case, if we do not assume world sheet picture, then only gauge invariance will be a hint. A building block is the open string vertex which is non-commutative but associative. However, known closed string vertices are non-associative and recovering them without worldsheet picture is quite nontrivial. 
So, conversely, seeking a geometric picture for generic $n$ may be important. We hope to report on this issue elsewhere.


\section*{Acknowledgements}
The work of M.A.~was supported in part by 2013 SEIKEIKAI Grant.
The work of M.K.~was supported in part by JSPS KAKENHI Grant Number 25287049.
\appendix 
\def\thesection{Appendix~\Alph{section}}
\renewcommand{\theequation}{\Alph{section}.\arabic{equation}}
\def\thesection{Appendix~\Alph{section}}
\setcounter{equation}{0}
\setcounter{figure}{0}
\def\thesection{Appendix~\Alph{section}}
\section{$\!\!$Basic properties of open string state space}
\label{app0}
\setcounter{equation}{0}
We give some of the basic properties of open string state space ${\cal H}(p)$.
It is spanned by states of the form 
\begin{equation}
\ket{f} = \alpha_{l_1}^{\mu_1} \cdots \alpha_{l_q}^{\mu_q}
c_{-m_1}\cdots c_{-m_r} b_{-n_1}\cdots b_{-n_s} \ket{\downarrow,0,p}
\end{equation}
with $l_i\ge 1$,  $m_i\ge 0$ and $n_i\ge 1$.   
Here, $\alpha_{l}^{\mu}$ is the matter oscillator, and $c_{m}$ and $b_{n}$ are the string worldsheet ghost and anti-ghost modes.
Note that $\ket{\downarrow,0,p} = \ket{0,p}\otimes c_1^{(i)}\ket{0}_{\rm g}$
where $\ket{0,p}(= e^{i p_\mu x^\mu} \ket{0}_{\rm M})$ is the momentum eigenstate annihilated by $\alpha_l^\mu$ for $l \ge 1$ 
and $c_1^{(i)}\ket{0}_{\rm g}(=\ket{\downarrow})$ is  
the ghost ground state annihilated by $c_m$ and $b_m$ for $m\ge 1$.

The (anti-)commutation relations among $\alpha_{n}^\mu$, $c_{n}$ and $b_{n}$ are 
\begin{equation}
[\alpha_{m}^\mu, \alpha_{n}^\nu ] =m\eta^{\mu\nu}\delta_{m+n,0},
\qquad \{b_{m},c_n \}= \delta_{m+n,0},
\quad \{b_m,b_n\}=\{c_m,c_n\}=0.
\end{equation}
The BRST operator is given by 
$Q=\tilde{Q}+c_0 L_0+b_0 M$ where  
\begin{equation}
\tilde{Q}= \sum_{n\ne 0} c_{-n} L_n^{({\rm m})}  
-\frac{1}{2} \sum_{\parbox{12mm}{\tiny\rule{5pt}{0pt}$mn\ne 0$\\$m+n\ne0$}} (m-n)\,:c_{-m} c_{-n} b_{n+m}: 
,\qquad
M=-2\sum_{m=1}^\infty m c_{-m} c_m
\end{equation}
and $L_n^{({\rm m})}$ is the matter part of total Virasoro operator
$L_n= L_n^{({\rm m})} + L_n^{({\rm g})} $.
They are explicitly given by 
\begin{equation}
L_m^{({\rm m})} = \frac{1}{2} \sum_{n=-\infty}^{\infty} : \alpha_{m-n}^\mu \alpha_{\mu, n}:,
\qquad
L_m^{({\rm g})} =  \sum_{n=-\infty}^{\infty} (m-n): b_{m+n} c_{-n}: -\delta_{m,0} .
\end{equation}
In particular, 
\begin{equation}
L_0= \alpha' p^2 +N-1
\end{equation}
where $\alpha'$ is the slope parameter and 
\begin{equation}
N=\sum_{n=1}^\infty 
\left(\alpha^\mu_{-n}\alpha_{\mu, n}
+ n (c_{-n} b_{n}  + b_{-n} c_{n})
\right) 
\end{equation} 
counts the level of the state.
The commutation relations for the Virasoro operators are given by
\begin{equation}
[L_n, L_m]= (n-m) L_{n+m}  + \frac{D-26}{12} n(n^2-1)\delta_{n+m,0}.
\end{equation}
For the spacetime dimension $D=26$, $Q^2=0$ is satisfied.

\section{Projection operators for the $a$-gauges}
\label{app1}
We give two types of projection operators defined on the space 
${\cal F}_n^{\hat{g}=m}$
which are important in defining the $a$-gauges.

First operators are defined on the spaces $\tilde{\cal F}_n^{\hat{g}=\pm m}$ with $m\ge 0$ as 
\begin{equation}
{\cal P}_m=-\frac{1}{L_0}  \tilde{Q} {\cal M}^m {\cal W}_{m+1} \tilde{\cal Q},
\qquad
{\cal P}_{-m}=-\frac{1}{L_0}  \tilde{Q} {\cal W}_{m+1} \tilde{Q} {\cal M}^{m}.
\end{equation}
In particular, $1-{\cal P}_0$ extracts the states satisfying $\tilde{\cal Q}\ket{f}=0$.

Next is the set of operators ${\cal S}_k$ with integer $2k$. 
This operator ${\cal S}_k$ extracts the SU(1,1)-spin$=k$ part from a given state $\ket{f}$.
For a state $\ket{f^{\pm m}}\in {\cal F}_n^{\hat{g}=\pm m}$ ($m\ge 0$), ${\cal S}_k$ is explicitly given by
\begin{eqnarray}
{\cal S}_k &=& {\cal M}^{k+ \frac{m}{2}} {\cal W}_{2k} {\cal M}^{k-\frac{m}{2}} 
- {\cal M}^{k+1+ \frac{m}{2}} {\cal W}_{2k+2} {\cal M}^{k+1-\frac{m}{2}} 
\qquad (\mbox{on ${\cal F}_n^{m}$})
\\
{\cal S}_k &=& 
{\cal M}^{k- \frac{m}{2}} {\cal W}_{2k} {\cal M}^{k+\frac{m}{2}} 
- {\cal M}^{k+1- \frac{m}{2}} {\cal W}_{2k+2} {\cal M}^{k+1+\frac{m}{2}} 
\qquad (\mbox{on ${\cal F}_n^{-m}$})
\end{eqnarray} 
for each $k \in \{ \frac{m}{2} +\pmb{Z}_{\ge 0}  \}$, and $S_k=0$ for otherwise.
Note that the following relations are satisfied:
\begin{equation}
{\cal S}_k {\cal S}_{k'} =\delta_{kk'} {\cal S}_k ,\qquad
[{\cal S}_k, {\cal M}]=0,\qquad [{\cal S}_k, {\cal W}_{\pm m}]=0 
\mbox{~~(on ${\cal F}_n^{\pm m}$)}.
\end{equation}


\end{document}